\documentclass{proc}
\usepackage{epsfig}
\begin{document}

\title{Highlights from XMM-Newton}

\noindent {\bf Highlights from XMM-Newton}

\noindent Xavier Barcons. Instituto de F\'\i sica de Cantabria
 (CSIC-UC), 39005 Santander \hfill {\it barcons@ifca.unican.es}\\

\medskip

\noindent Ignacio Negueruela.
Departamento de F\'\i sica, Ingenier\'\i a de Sistemas y Teor\'\i a de
la Se\~nal,  
Universidad de Alicante, 03080 Alicante\hfill 
 {\it ignacio@dfists.ua.es}\\

\medskip

\section*{Abstract}

The launch of the {\it Chandra} (NASA) and {\it XMM-Newton} (ESA) X-ray
observatories in 1999 has revolutionized our view of the Universe, by
providing astrophysical information about many classes of sources with
unprecedent detail.  The high throughput of {\it XMM-Newton} makes it
the ideal instrument to provide low to moderate resolution
spectroscopy of faint and extended sources. After 3 years of
operations, {\it XMM-Newton} has observed all types of astronomical sources
and delivered very interesting results in many areas. In this review,
we highlight a few points where the contribution of {\it XMM-Newton}
has significantly furthered our knowledge of the energetic Universe.

\section*{Resumen}

El lanzamiento de los observatorios de rayos X {\it Chandra} (NASA) y
{\it XMM-Newton} (ESA) en 1999 ha revolucionado nuestra visi\'on del
Universo, al proporcionar informaci\'on astrof\'\i sica detallada sobre muchos
tipos de fuentes. La alta eficiencia de {\it XMM-Newton} lo convierte
en el instrumento ideal para obtener espectroscop\'\i a de
resoluci\'on baja y moderada de fuentes d\'ebiles y extensas.
Despu\'es de 3 a\~nos de operaciones, {\it XMM-Newton} ha observado
todo tipo de fuentes astron\'omicas y ha proporcionado resultados de
la mayor importancia en muchas \'areas.  En esta revisi\'on,
destacamos unos pocos aspectos donde la contribuci\'on de {\it
XMM-Newton} ha incrementado de forma significativa nuestro
conocimiento del Universo energ\'etico.

\section{Introduction}

X-ray Astronomy is at the very heart of the enormous progress that our
astrophysical knowledge of the Universe has undergone over the last
decades.  X-rays are emitted in a wide range of physical situations in
the Universe, invariably linked to the presence of hot gas and strong
gravitational fields.  These phenomena, largely unsuspected in the
early days of X-ray astronomy, are nowadays seen to be ubiquitous in a
variety of astronomical objects. X-ray observations of previously
known objects have often revealed phenomena (e.g., the presence of an
accreting black hole) just not seen at other wavelengths.  In
addition, new sources serendipitously discovered in X-ray surveys have
uncovered new classes of objects, totally inconspicuous at optical
wavelengths.  One datum that illustrates the impact of X-ray
observations in Astronomy is that 20\% of the papers published by the 3
main Astrophysics journals in 2002 (Astrophysical Journal, Monthly
Notices of the Royal Astronomical Society and Astronomy \&
Astrophysics) contain the ``X-ray'' keyword in the abstract. The 2002
Nobel Prize in Physics given to Riccardo Giacconi (father and leader
of X-ray Astronomy), is also a sign of the good health reached by this
discipline, as it is one of the very few granted to astronomers.

\subsection{Some history}

X-ray Astronomy began in 1962, when a rocket flown to detect the Moon
in scattered X-rays from the Sun, discovered instead the first
extra-solar X-ray source (Sco X-1) and the cosmic X-ray background
(Giacconi et al. 1962).  This already showed that the Universe out
there could be radically more energetic than what was known before, as
Sco X-1 had a much larger X-ray to optical flux ratio than the
Sun. More rockets followed this pioneering discovery, but there was a  basic
limitation in the amount of observing time available for observations
in a single flight.

{\it UHURU} (which means {\it freedom} in swahili) was the first
orbiting X-ray observatory. It was launched from Italy's San Marco
station in Kenya on the 12th of December of 1970, coinciding with the
independence day of this country (thence the name). The {\it UHURU}
payload weighed less than 60 kg.  It scanned the sky for over 2 years
and produced the first all-sky catalogue of X-ray sources, containing
a few hundred of them.  Subsequent work demonstrated that outside the
galactic plane, most of the sources were Active Galactic Nuclei (AGN)
and clusters of galaxies.  In the Galactic plane, {\it UHURU} found
X-ray binaries, supernova remnants   and other diffuse structures.  This
rough picture (but with varying fractions of objects) still describes
to zero-th order the present knowledge of the X-ray Universe.

This early generation of X-ray orbiting observatories (which also
included {\it HEAO-1} and more recently {\it Ginga}, Japanese word for
Milky Way) were equipped with a mechanical collimator as the only means
to limit the field of view covered by the (usually gas-filled
proportional counter) detectors, without any further optics.  This
provided a very rough angular resolution, of the order of several
degrees, which ultimately limited the sensitivity of the observatory
because of confusion of fainter sources. 

Proper imaging X-ray optics was featured for soft X-rays (below 4.5
keV) with the {\it Einstein Observatory} first.  With an angular
resolution of a couple of arc minutes, {\it Einstein} discovered many
new X-ray sources (including many at cosmological distances), showing
that the word {\it experiment} was in the way of being replaced by
{\it observatory} in X-ray missions. {\it Einstein} made it possible
to discover new classes of sources by their X-ray emission, resolved
the spatial structure of the intracluster gas in galaxy clusters and
produced large catalogues of X-ray sources that were subsequently used
to conduct detailed astrophysical studies.  It also showed, for the
first time, that a significant fraction of stars are X-ray
emitters. {\it ROSAT}, in some sense a successor to {\it Einstein},
deepened our knowledge of the X-ray Universe, by conducting first an
all-sky survey (this yielded the, so far, largest catalogue of X-ray
sources, exceeding 50000), and for almost a decade, targeted
observations of a full range of astronomical objects, from comets to
distant QSOs.

Focusing X-ray photons is not an easy task. In a standard reflecting
optical telescope, photons are directed almost normally to the
reflector's surface and collected in the focal plane.  If the same
setup is used for X-rays, they end up either transmitted or absorbed,
but never reflected.  Grazing incidence is needed to achieve total
reflection for X-rays, the maximum angle with respect to the surface
being of a few degrees for a photon of $\sim 1\, {\rm keV}$.
Reflection becomes more difficult at higher photon energies, as the
angle for total reflection becomes smaller and the mirrors need to be
placed almost parallel to the optical axis.  It is easy to see that
the effective area that a photon sees is very small with a mirror
placed in grazing incidence, and therefore several (or many) mirrors
are nested one inside each other to increase the collecting area.  It
is also clear that an X-ray focusing telescope of higher energy
photons will have less effective area and longer focal length than a
similar one focusing soft X-ray photons.

The start of the nineties brought X-ray observatories
able to produce images with photons of up to 10 keV, such as the
Japanese {\it ASCA} were launched.  {\it ASCA} also opened the door to
a more modern type of X-ray detector (the {\it Charge Coupled Device}
or CCD) that ended up burying the proportional counters. CCD-based
detectors deliver an order of magnitude better spectral resolution for
single photons than proportional counters.  At the energy of the Fe
K$\alpha$ line (6.4-6.7 keV), the resolution of $\sim 50$ achieved by
these detectors enables a detailed study of the physical environment
where it is produced.

Other X-ray observatories have been launched during the 1990's,
including {\it BeppoSAX} (which also featured higher energy detectors
sensitive to many tens of keV) and {\it Rossi X-ray Timing Explorer}
(RXTE) which is tailored to timing analysis of bright sources.
However, at the turn of the new millenium, both {\it NASA} and {\it
ESA} decided to launch their respective large X-ray observatories:
{\it Chandra} (launched July 1999) and {\it XMM-Newton} (launched
December 1999).  The coincidence of operations between both missions
(being just by chance, as {\it Chandra} was over-delayed for several
years) has brought what can be called the {\it golden age} of X-ray
Astronomy.  By virtue of their respective designs {\it Chandra} is
mostly an imaging observatory, while {\it XMM-Newton} has its major
capabilities at spectroscopy.  Having both of them operating at the
same time has opened a huge window to X-ray astrophysics, whose
dimensions are just beginning to be realized after the first few years
of operations.

%\subsection{The physics of X-ray Astronomy}

\section{XMM-Newton}

{\it XMM-Newton} is an X-ray observatory launched and operated by the
European Space Agency (ESA), with instruments contributed and funded
by ESA member states and NASA (USA).  {\it XMM-Newton} was
successfully launched by an Ariane 5 on the 10th of December of
1999. Its orbit is highly eccentric with a period of $\sim 2$ days.
Although the original {\it XMM-Newton} project was approved for 2
years of scientific operations, all systems and payloads are designed
to last for $\sim 10$ years.  Specifically, fuel (hydrazine) is
currently thought to last for more than that.  In fact, an extension
of the {\it XMM-Newton} operations has already been approved to 2006
with a further 2 year period being revised by mid 2003. For comparison
with early missions, the full {\it XMM-Newton} satellite weighs more
than 3000 kg. Currently, the {\it XMM-Newton} spacecraft is operated
by ESOC at Darmstadt (Germany), but the instruments and science
operations in general are managed from VILSPA (Villafranca del
Castillo, Spain).

\begin{figure}
\label{XMM}
\caption{Scheme of the grazing incidence optics on the {\it
XMM-Newton} X-ray telescopes (from ESA)}
\epsfig{file=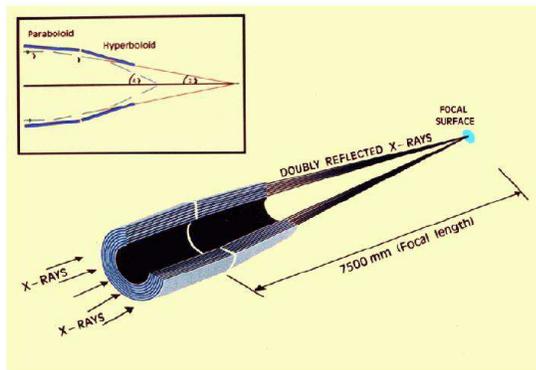,width=6cm,angle=270}
\end{figure}

{\it XMM-Newton} consists of 3 co-aligned grazing incidence X-ray
telescopes (Jansen et al. 2001), with angular resolution of 4-6 arcsec
(FWHM) and 13-15 arcsec (Half Energy Width).  Each X-ray telescope
nests 58 grazing incidence mirror pairs, with a total collecting area
of $\sim 4000\, {\rm cm}^{-2}$ (see Fig~\ref{XMM} for a schematic of
the X-ray optics in the XMM-Newton X-ray telescopes). One of these
telescopes focuses all X-rays into a single spectroscopic imaging
detector (EPIC-pn, Str\"uder et al. 2001), which has a pixel size of 4
arcsec and covers a field of view of about 30 arcmin in diameter.
About half of the X-rays focused by the other two telescopes go
undispersed to similar imaging spectrographs called EPIC-MOS (Turner
et al. 2001) . The EPIC instruments can measure the energy of
individual photons with a resolution of $\sim 20-50$. EPIC is
sensitive to photons from 0.2 to 12 keV and can detect X-ray sources
as faint as $\sim 10^{-15}\, {\rm erg}\, {\rm cm}^{-2}\, {\rm s}^{-1}$
in the 0.5-2 keV band in a few tens of kiloseconds (ks).  Its main
limiting sensitivity in soft X-rays is caused by confusion, due to the
modest angular resolution of the X-ray telescopes. At higher photon
energies the main limiting factor is photon counting, with an expected
confusion limit in the 2-10 keV band reached in a several Ms exposure
only.

The two X-ray telescopes that focus the X-rays into the EPIC-MOS
detectors are equipped with difraction gratings working by reflection.
These disperse the remaining X-rays according to their wavelength and
these are recorded in a further chain of CCD detectors.  This
wavelength-dispersive instrument is called the Reflection Grating
Spectrometer (RGS) and delivers a spectral resolution $\sim 200$ in
first order dispersion (i.e., $\sim 0.06\, $\AA)  in the soft X-ray
domain (5-35 \AA, or 0.3-2.4 keV).  The sensitivity of the RGS can be
viewed in terms of a source with 0.5-2 keV flux of $\sim 10^{-11}\, {\rm
erg}\, {\rm cm}^{-2}\, {\rm s}^{-1}$ producing peak
S/N$\sim 10$ per spectral resolution element in 100 ks. It is
therefore an instrument designed to observe bright sources with
relatively high spectral resolution. The moderate angular resolution
of the XMM-Newton mirrors makes it possible to obtain integrated
moderate resolution of {\it extended} sources, a capability that is
producing extremely interesting results (see later).

Besides that, {\it XMM-Newton} has a co-aligned optical/UV telescope (the
Optical Monitor - OM) which operates in the range from 1600 to 6600$\,
{\rm \AA}$. At its focal plane the OM has a photon counting device
(Mason et al. 2001), which delivers time-resolved information for the
detected photons. The field of view of the OM is 17 arcmin, with a
point-spread-function of FWHM between 1.3 and 2.5 arcsec, with a
limiting sensitivity of 23.5$^{\rm mag}$ for 1000 sec integration.
The OM is equipped with a number of UV and optical filters, plus low
resolution UV and optical grisms, whose calibration is being conducted
at the time of writing this report. All three {\it XMM-Newton}
instruments (EPIC, RGS and OM) are simultaneously operated.

The ground segment operations of {\it XMM-Newton} are conducted by the SOC at
Villafranca del Castillo (Spain) with the assistance of an ESA
member-state funded consortium called {\it Survey Science Centre} (SSC).
The SSC tasks include the development of Science Analysis Software tasks
in collaboration with the SOC; the pipeline processing of all {\it XMM-Newton}
data and the identification of the {\it serendipitous} sources
discovered by {\it XMM-Newton}.  This last task is specially demanding, as it
is likely that {\it XMM-Newton} will find about 50000 new X-ray sources per
year. Identifying and cataloguing these data will constitute a major
legacy of {\it XMM-Newton} (see Watson et al. 2001). 

ESA has already released the first version of the {\it XMM-Newton} catalogue
provided by the SSC, which contains over 30000 good-quality X-ray
sources. Some of these sources are identified thanks to the extensive
archival search conducted by the SSC.  At the time of writing this
report the SSC has imaged in various optical filters a large fraction of
fields that contain the catalogued sources. Besides that, well
over 500 of these X-ray sources have been spectroscopically identified.
All this optical imaging and spectroscopic information will be included
in future versions of the {\it XMM-Newton} catalogue, delivering an extremely
powerful tool to investigate the X-ray sky.

\section{Normal stars}

One of the areas where the impact of {\it XMM-Newton} has been
strongest is the study of X-ray emission from ``normal'' stars. The
possibility that stars without compact companions could be X-ray
emitters had not been initially foreseen. Though some normal stars had
been detected by previous experiments (e.g., Topka et al. 1978), only
after the observations by {\it Einstein} was it fully realised that
many solar-type stars emitted soft X-rays (Vaiana et al. 1981).
Equally surprising was the discovery that hot OB stars also appeared
to be substantial soft X-ray sources.

The X-ray emission from low-mass stars was quickly interpreted in
terms of coronal activity, similar to that observed in the Sun, but on
a much larger scale. Many M-type stars were found to display
relatively strong X-ray emission, which correlated with other
indicators of activity, such as emission in H$\alpha$ or the Ca\,{\sc
ii} doublet (Noyes et al.\ 1984; Vilhu 1984, Fleming et al.\ 1989).

It was also found that there was a strong correlation between age
and X-ray activity. In this sense, the all-sky survey by {\it ROSAT} greatly
changed the previous view on the evolution of low-mass stars towards
the main sequence. It was seen that many stars not displaying the
typical characteristics of T Tauri stars could be identified as very
young objects because of their X-ray emission (e.g., Neuh\"auser
1997). This lead to the
discovery of whole new associations of young stars. As previous
missions lacked spectral resolution in the soft X-ray range, {\it
XMM-Newton} and {\it Chandra} offer the possibility of obtaining, for
the first time, information about the physical causes of this emission.

\subsection{Coronal activity}

Coronal activity in chromospherically active low-mass stars is thought
to result in X-ray emission through mechanisms similar to those
observed in the Sun. Observations with {\it XMM-Newton} provide
spectra showing lines from a large variety of elements (O, Ne, Mg, Fe,
N, etc), many of them in at least two different ionisation states.
From them, coronal abundances, which are in many cases very different
from the stellar atmospheric abundances, are derived.  In the Sun,
elements with low first ionisation potential are overabundant with
respect to their photospheric abundances, while elements with high
first ionisation potential are not. In a survey of RS CVn binaries
(tidally locked binaries containing a highly active slightly evolved
G-K star), Audard et al.\ (2003) find that the most active stars
display a behaviour opposite to that seen in the Sun, suggesting that
fractionation mechanisms should be revised.  They indicate that this
opposite effect is seen in all stars with high coronal activity, while
stars with lower activity seem to present an effect similar to the
Sun.

Simultaneous observations of $\sigma$ Gem, an RS CVn binary with a
K1III primary, with {\it XMM-Newton} and the VLA have allowed the
discovery of a correlation between the radio luminosity of a large
flare and the time derivative of the X-ray luminosity (G\"udel et
al.\ 2002). This
relationship is observed in solar flares, where it is known as the
Neupert effect, and supports flare mechanisms causing chromospheric
evaporation.

The presence of X-ray emission from young (pre-main-sequence)
stars is now recognised as a widespread phenomenon (Feigelson \&
Montmerle 1999). It is assumed to be a consequence of the same
magneto-centrifugal processes that drive the outflows and winds
associated with classical T Tauri stars (e.g., Shu et
al.\ 1994). Large-scale surveys with {\it XMM-Newton} will allow an
understanding of these phenomena. As a first example, Favata et
al.\ (2003) have observed the L1551 star forming complex. They find
that the characteristics (both temporal evolution and spectrum) of
classical T Tauri stars and weak-lined T Tauri stars are very
different, suggesting that while weak-lined T Tauri stars show an
enhanced version of the coronal activity seen at older ages, classical
T Tauri stars have a different X-ray emission mechanism, related to
their accretion disks.

\subsection{Massive stars}
Although emission from low mass stars can be understood in terms of
coronal activity, the existence of relatively soft X-ray emission from
massive stars came as a bit of a surprise (Harnden et
al.\ 1979). Models trying to explain it invoke hydrodynamic shocks
resulting from intrinsic instabilities in their radiatively driven
winds (see Feldmeier et al.\ 1997). {\it XMM-Newton} RGS  spectra have
allowed the exploration of physical conditions in the regions where
the emission is produced. Observations of the O4Ief star $\zeta$
Puppis have provided confirmation that the X-ray emission consists
mostly of broad emission lines from H-like and He-like
charge states of N, O, Mg and Si, as well as Ne-like ions of Fe (Kahn
et al.\ 2001). In this object with a very strong radiative wind, X-ray
emission reaching us seems to be produced at large distances from the
star.

\begin{figure}
\label{zpup}
\caption{Top panel: The RGS spectrum of $\tau$ Sco, rebinned by a factor 3,
compared to the EPIC MOS spectrum of the same star. The impressive
spectral resolution of the RGS stands out. Prominent lines are labeled
with the emitting ions (from Mewe et al. 2003).
 Bottom panel: The RGS spectrum of $\zeta$ Pup from Kahn et al. (2001).
The lines are clearly much broader than in $\tau$ Sco.}  
\epsfig{file=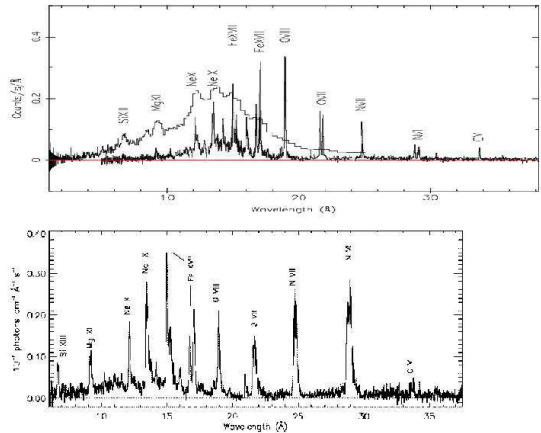,width=6cm,angle=270}
\end{figure}

 Conversely, the X-ray spectrum of the B0.2V MK standard $\tau$
Scorpii is not very consistent with standard models. Emission lines
in its  {\it XMM-Newton} RGS  spectrum are as narrow as the
instrumental profile, suggesting that they are produced much closer to
the star (Mewe et al.\ 2003). X-ray emission shows a hot component
corresponding to a temperature in excess of $20\times10^6\:$K. This is
supported by a {\it Chandra} observation of the same star by Cohen et
al.\ (2003), who suggest that the spectrum of $\tau$ Scorpii is
intermediate between those of O-type stars and coronal emitters. As
there is rather strong evidence against the possibility of a cool
companion for $\tau$ Scorpii, a model in which dense clumps in the
wind stall and decelerate, perhaps falling back onto the stellar
surface, such as that presented by Howk et al.\ (2000), is
preferred. The hard X-rays would then originate in  bow shocks around
those clumps.

In some cases, emission from relatively hot stars seems to be due to
late-type companions. Simultaneous observations of Castor with {\it
XMM-Newton} and {\it Chandra} allow the resolution of its three
components, all of which are close binaries. Castor C is a pair of M1V
stars, both of which appear to be moderately active (Stelzer et
al.\ 2002). Both Castor A and
Castor B contain an A-type star and a low-mass companion, but their
emission appears typical of coronal activity in the low mass stars
(Stelzer \& Burwitz 2003).

Another interesting result obtained with {\it XMM-Newton} is the lack
of X-ray emission from C-rich Wolf-Rayet stars. Oskinova et al.\ (2003)
failed to detect WR 114 with {\it XMM-Newton}, implying a lower limit
on its X-ray luminosity
$L_{{\rm x}}/{L_{\rm bol}}\leq4\times10^{-9}$.
Until now, no single WC star has been detected in X-rays, while
several WN star have. An example is WR 110, detected with {\it
XMM-Newton} by Skinner et al.\ (2002). Surprisingly, the X-ray spectrum
of this very hot star also shows a hard component which cannot be
explained with current models.

In some cases, a relatively hard X-ray spectrum is produced when the
winds of two massive stars forming a close binary collide. This is the
case of HD 93403 (O5.5I + O7V), where clear orbital modulation has
been detected with {\it XMM-Newton} (Rauw et al.\ 2002).

\section{Accreting binaries}

X-ray binaries (XRBs) are systems in which high energy radiation is
emitted
as a consequence of the accretion of material from a donor star (in
most cases, a hydrogen-burning ``normal'' star) onto the surface of a
compact object (see Lewin et al.\ 1995, for extended reviews). Binary
systems in which the
accreting compact object
is a white dwarf are generally considered as a separate subclass,
known as Cataclysmic Variables, though there are no strong physical
reasons for this separation.

XRBs emit most of the X-ray flux detected from normal
galaxies, including the Milky Way. They have been discovered in large
numbers by
previous satellites and now a sufficiently large sample exists for
population studies to be statistically significant (e.g., Helfand \&
Moran 2001). {\it XMM-Newton}
offers several potentialities for their study. On the one
hand, it opens up the opportunity of obtaining spectra of Galactic
sources with moderate resolution and high signal-to-noise ratio. On
the other hand, because of its large collecting area and moderately
high spatial resolution, it provides an excellent opportunity for
studying the populations of nearby galaxies. Last but not least, {\it
  XMM-Newton}  offers the possibility of obtaining X-ray spectra of
faint Galactic sources, opening the door to studies of low-luminosity
accreting binaries.

XRBs are generally divided into two main subclasses, high
mass (HMXBs) and low mass (LMXBs), depending on the nature of the
donor star (either
an OB star or G$-$M spectral type). Additionally, depending on the
temporal behaviour, they can be divided in persistent and transient
sources. LMXBs containing an accreting neutron star
can be either persistent or transient, but systems with black holes
are generally transient.

\subsection{Low mass X-ray binaries}
Classical low mass X-ray binaries (LMXBs) consist of a neutron star with a
moderately strong magnetic field ($\sim 10^{8}\:$G) accreting from a
low-mass star. The light-curves of these objects are rather complex
(displaying temporal features such as eclipses and dips). There is a
large variety of behaviours among LMXBs, likely due to the complexity
of their accretion geometries, which has prevented the creation of a
unified model. As a matter of fact, while all LMXBs observed with {\it
ASCA} could be well fit with a two component model, a blackbody
representing the accretion disk and an extended Comptonising region
(Church \&
Balucinska-Church 2001), this model failed to fit observations of
several LMXBs with {\it BeppoSAX} (e.g., Oosterbroek et al.\ 2001).

Many LMXBs also display bursts, very rapid
increases in the X-ray flux followed by exponential declines
(typically lasting seconds to minutes). The bursts are interpreted as
thermonuclear explosions on the stellar surface, after
material has accumulated due to accretion (Lewin et al.\ 1993). During
this bursts, X-ray emission from the vicinity of the neutron star
dominates that coming from the disk.

{\it XMM-Newton} has allowed the study of the accretion environments
of low mass X-ray binaries , with the discovery of complex narrow
absorption features in their X-ray spectra, corresponding to H-like
and He-like ions of O and Fe, and likely other elements (Cottam et
al.\ 2001; Sidoli et al.\ 2001; Parmar et al.\ 2002). For example,
comparing the number of absorbed photons at the O\,{\sc vii} and
O\,{\sc viii} edges with the number of photons emitted in the O\,{\sc
viii} Ly$\alpha$ and O\,{\sc vii} He-like complexes, Cottam et
al.\ (2001) argue that absorbing material must be aligned with the
accretion disk and extend along our line-of-sight.

Jimenez-Garate et al.\ (2002) are able to derive elemental abundance
ratios
from recombination emission lines seen in the spectrum of the LMXB Her
X-1. They derive very high N/O ratios, which they interpret as proof
of strong interactions during the formation history of this binary.

\subsection{Soft X-ray transients}

LMXBs containing a black hole (and a few transient sources containing
neutron stars) are generally termed Soft X-ray
transients (SXTs) or X-ray novae. This is because they display very
strong X-ray and optical outbursts (generally also accompanied by
radio emission) separated by long periods of quiescence (see Tanaka \&
Shibazaki 1996). They are
``soft'' in the sense that their outburst spectra are dominated by a
soft blackbody component at $\sim1\:$keV. The X-ray spectra of SXTs are
very
variable and characteristic low/hard and high/soft states have been
identified. The geometry of the accretion flow and the source of soft
photons during the outbursts is currently debated. In order to explain
their long quiescent states optically thin advection-dominated
accretion flows (ADAFs) have been invoked (Narayan et al.\ 1996; Esin et
al.\ 1998). Some authors have argued that the very low X-ray fluxes
during quiescence could be due simply to coronal activity from the
donor.

The large collecting area of {\it XMM-Newton} allows the
observation of very faint sources in quiescence. The SXT GU Mus was
observed in quiescence as a very faint source. The X-ray flux was
characterised by a power law, ruling out coronal activity and
apparently supporting ADAF models (Sutaria et al.\ 2002). Similar
conclusions were drawn from an observation of the SXT GRO J1655$-$40
(Hameury et al.\ 2003). However,
observations of the black hole candidate and micro quasar GRS
1758$-$258 found a very soft spectrum in the off state, against the
predictions of simple ADAF models (Miller et al.\ 2002).

Another source, XTE
J1650$-$500, was observed in the very high state by Miller et
al.\ (2003a). Its spectrum displayed broad Fe K$\alpha$ lines,
whose shape suggests that rotational energy is being extracted from a
spinning black hole.

\subsection{Cataclysmic variables}
Cataclysmic variables (CVs) are systems containing a white dwarf
accreting from a low-mass hydrogen-burning star (e.g., Sion 1999).
As such, CVs may manifest themselves under the guise of classical
novae, dwarf novae, recurrent novae, nova-likes and similar objects
(Patterson 1984). X-ray emission is
generally detected from those CVs in which the white dwarf exhibits a
strong magnetic field, the polars or AM Her stars, or
a moderate magnetic field, the intermediate polars or DQ Her stars
(Patterson 1994). The interest of CVs stems from the fact that they
allow a detailed study of accretion processes, with wide applications
in several astrophysical contexts.

As the X-ray spectra of CVs are not as hard as those of accreting
neutron stars and their luminosities are also not very high, sensitive
instruments are needed to detect their emission. {\em ROSAT} surveys
resulted in the discovery of large numbers of new
faint CVs and it is expected that {\it XMM-Newton} will provide
spectral information for a large number of sources, making statistical
studies possible. Such work is already in progress, with large numbers
of CVs having been observed (e.g., Ramsay \& Cropper 2002; Ramsay \&
Cropper 2003), allowing determination of their basic parameters, such
as white dwarf mass and mass transfer rate.

Of particular interest is the observation of the recent nova V2487 Oph
1998. This classical nova was detected with signatures of an accreting
white dwarf less than three years after its nova outbursts (Hernanz \&
Sala 2002). This detection suggests that the system
was back to its standard configuration after the phenomenal
thermonuclear explosion which originated the nova outburst.

\subsection{Supersoft sources}
Discovered originally in the LMC, supersoft X-ray sources (SSSs) display
X-ray spectra peaking at energies much lower than traditional X-ray
binaries ($\sim 40\:$eV). Because of this, they are heavily affected
by interstellar absorption and unlikely to be detected in the Milky
Way, unless they are very close. They are generally interpreted as
white dwarfs accreting from a more massive hydrogen-burning donor. The
mass transfer is therefore unstable and proceeds on the thermal
timescale (van den Heuvel et al.\ 1992). Because of their soft spectra,
they have only been studied in any detail with {\it ROSAT}. {\it
XMM-Newton} provides now the opportunity to observe their spectra.

The first RGS spectrum of an SSS, CAL 83 in the LMC, proved that the
X-ray emission originated from the photosphere of a very hot white
dwarf by showing a rich spectrum of absorption complexes due to
several elements (Paerels et al.\ 2001b). Even though this spectrum is
qualitatively similar to expectations from white dwarf atmosphere
models, RGS observations of the
Galactic SSS MR Vel reveal a wealth of
emission lines displaying  P-Cygni profiles, indicative of a
strong wind from the white dwarf's surface, which cannot be reproduced
with current models (Motch et al.\ 2002). Also interesting is the
likely detection of Doppler shifts due to orbital motion in the
X-ray emission lines of MR Vel, probably the first detection of
orbital motion at high energies (Motch et al.\ 2002).

In {\it XMM-Newton} pointings, Osborne et al.\ (2001) have found a
transient SSS close to the nucleus
of M31 displaying an 865-s periodicity. King et al.\ (2002) interpret
this as the spin period of a white dwarf and hence argue (see also
Schenker et al.\ 2002) that CVs are descendants from SSSs.

\section{Supernova Remnants}
Supernova explosions have a very profound impact on the Interstellar
Medium (ISM), both as sources of mechanical energy and heavy
elements. Supernova Remnants (SNRs) provide information about these
issues and can be observed as high-energy sources. The interaction of
high velocity ejecta with the ISM generates very high temperatures,
giving rise to line emission from heavy elements. By studying the
spatial distribution of elements, we can gain insight into
nucleosynthesis in the progenitor and the geometry of the explosions.
SNRs are also possible sites of cosmic ray acceleration.

The characteristics of the RGS make it possible to obtain spectral
information from relatively extended sources. RGS spectra of SNRs
allow the study of physical conditions all over the remnant (e.g.,
Rasmussen et al.\ 2001).
By measuring the mass, temperature and bulk velocity of material, it
is possible to obtain information about the supernova explosion.
Using {\it XMM-Newton} observations of the young SNR Cas A, Willingale
et al.\ (2002) have derived kinematic information and abundance
ratios for the supernova ejecta. From analysis of those data, they
find evidence for beaming in the supernova explosion, suggesting that
most of the material was ejected in two jets (Willingale et
al.\ 2003). They also suggest that the progenitor was a very massive
Wolf-Rayet star.

The compact object born in the Cas A supernova explosion has been
identified with a point-like {\it Chandra} source, CXO J232327.8+584842. A
long pointing with {\it XMM-Newton}, however, has failed to detect the
X-ray pulsations that would be expected from a neutron star
(Mereghetti et al.\ 2002a). The
spectral properties of this source are also difficult to interpret in
terms of a rapidly spinning neutron star.

No pulsations have been
detected either from the SNR G21.5$-$0.9, which in many aspects
resembles the Crab SNR, where a central pulsar is injecting high
energy electrons (La Palombara \&
Mereghetti 2002). In this SNR, the X-ray spectrum of the nebula
softens as one moves out from the centre, reflecting the energy losses
of the electrons
due to synchrotron radiation, as they diffuse out (Warwick et al.\ 2001).
Similar conclusions are reached for the SNRs G0.9+0.1 (Porquet et
al.\ 2003) and 3C~58 (Bocchino et al. 2001). While the central compact
object of G0.9+0.1 has been detected with Chandra (CXOU
J174722.8$-$280915), no obvious emission from a central object is seen
in 3C~58.

The SNR IC~443 is specially interesting, because it is interacting
with a dense molecular cloud rather than with the low density
ISM. {\it XMM-Newton} has resolved several discrete hard X-ray sources,
which could be fragments of the SNR
interacting with the dense cloud (Bocchino \& Bykov 2003).

\section{Neutron stars}
\subsection{Young neutron stars in supernova remnants}
The previous generation of imaging X-ray satellites (mainly {\it
ROSAT} and {\it ASCA}) have detected a large population of young
neutron stars which appear as rotation powered X-ray pulsars. X-ray
emission in these systems may arise from a variety of physical
processes. In many of them, non-thermal emission from relativistic
particles accelerated in the pulsar magnetic field shows a power law
spectrum over a broad energy range.

The large collecting area of {\it XMM-Newton} has provided for the
first time spectral information for many of these weak X-ray
sources. Its spatial resolution has also been necessary to separate
them from their surrounding supernova remnants (SNRs). Preliminary
results have been advanced by Becker \& Aschenbach (2002), and the
publication of many interesting results is expected in the near future.

\subsection{Millisecond pulsars}
Millisecond pulsars were first detected as radio sources (see Lorimer
2001) and only detected as X-ray pulsars with {\it ROSAT} (Becker \&
Tr\"umper 1993). It is generally believed that millisecond pulsars are
``recycled'' pulsars, which have been spun up to their present high
rotational velocities by accretion in a low mass X-ray binary and
therefore objects of the highest interest for our understanding of
binary evolution (Bhattacharya \& van den Heuvel 1991).

Because of this, the recent detections of millisecond pulsars in
accreting binaries (Wijnands \& Van der Klis 1998) has sparked
enormous interest. {\it XMM-Newton} observations of two of the
currently four known accreting millisecond pulsars have been
reported. The first accreting millisecond pulsar SAX J1808.4$-$365 has
been detected in quiescence (Campana et al.\ 2002) and at a slightly
higher luminosity (Wijnands 2003). In both cases, its spectrum was
dominated by a power law, incompatible with thermal emission from the
cooling surface of the neutron star (heated by accretion). A much
better spectrum was obtained for the 2.3-ms accreting millisecond
pulsar XTE J1751-305 (Miller et al.\ 2003b). However, no spectral
features were detected.

The apparently isolated 4.86-ms pulsar PSR J0030+0451 was detected by
{\it XMM-Newton}, which was able to measure its pulsed fraction. Its
spectrum could be fitted by a two component model (either blackbody
plus power law or two blackbodies) or a broken power law (Becker \&
Aschenbach 2002). Current models accounting for the spectra of
millisecond pulsars predict that PSR J0030+0451 should be detectable
at optical wavelengths. However, a very deep search in its {\it
XMM-Newton} error circle conducted with the VLT has failed to detect
any source down to B=27.3, casting into doubt such models (Koptsevich
et al. 2003).

\subsection{Thermal emission from neutron star atmospheres}
Observations of isolated neutron stars (or neutron stars in quiescent
X-ray binaries) are extremely important in fundamental physics, as thermal
emission from the surface of a neutron star carries signatures
of its gravitational field, which may be used to infer its mass and
radius. Detection of absorption lines corresponding to elements on the
neutron star atmosphere and measurement of their gravitational
redshift would provide rather accurate data (e.g., \"Ozel \& Psaltis
2003).From the gravitational redshift at the surface of the neutron
star, the ratio between its mass and radius may be measured, providing a
very strong constraint on neutron star models. Such models give Physics an
experimentally testable handle on properties of matter at (supra-)nuclear
densities.

Unfortunately, observations so far have not been very successful at
detecting these atmospheric features.The first {\it XMM-Newton}
observation of an isolated neutron star (RX J0720.4-3125) yielded no
spectral features on top of the black body continuum in a 62.5 ks
exposure (Paerels et al. 2001a).  Also, the isolated neutron star RX
J1856.5$-$3754 was observed for 57 ks with {\it XMM-Newton} without
revealing any spectral feature. An extremely long (505 ks) exposure
was then performed with {\it Chandra}, also failing to detect any
feature (Burwitz et al.\ 2003). This result is against predictions by
most current model atmospheres for neutron stars.

The spectrum of RBS 1223 does show a very broad absorption line, but this
is interpreted as a cyclotron feature caused by electrons moving in
its very large magnetic field of $2-6\times 10^{13}\, {\rm Gauss}$ (Haberl
et al. 2003). No lines have been seen in the spectra of several
other isolated neutron stars observed with {\it Chandra} or the
anomalous X-ray pulsar 1E1048.1$-$5937 observed with {\it XMM-Newton}
(Tiengo et al.\ 2002). The only secure detection is that of two
absorption features, discovered with {\it Chandra}, in the isolated
neutron star 1E 1207$-$5209. {\it XMM-Newton} observations have shown
a phase dependence in at least one of these features (Mereghetti et
al.\ 2002b), but their identification is still not clear.

Absorption features
from the surface of a neutron star have likely been identified during
a burst from the LMXB EXO 0748$-$676 (Cottam et al.\ 2002). If the
identification of the observed features with Fe\,{\sc xxvi} and {\sc
xxv} and  O\,{\sc viii} lines is correct, they imply a $z=0.35$
gravitational redshift.

\section{Normal galaxies}
As indicated above, X-ray emission from normal galaxies arises from
their populations of X-ray binaries. By comparing the populations of
different galaxies, we are able to obtain information about their star
forming histories (e.g., Grim et al.\ 2003). In order to resolve the
populations, imaging capabilities are necessary. Because of this,
thorough exploration of the most nearby galaxies, the Magellanic
Clouds, started only with {\it Einstein} and flourished with {\it
ROSAT} (e.g., Sasaki et al.\ 2000; Haberl et al.\ 2000). {\it ROSAT}
also allowed the study of other nearby galaxies such as M31 (Supper et
al.\ 2001) or M33 (Haberl \& Pietsch 2001).

The potentialities of {\it XMM-Newton} for this sort of study are
enormous. Deep pointings of the Magellanic Clouds result in the
detection of a wealth of X-ray sources, many of which are accreting
binaries. The sensitivity of {\it XMM-Newton} permits the detection of
effectively any active XRB. A pointing at the North of the LMC
resulted in the detection of 150 discrete sources, among which a newly
discovered HMXB, a new likely SSS and several new SNRs (Haberl et
al.\ 2003). A pointing to the SMC resulted in the detection of the
pulse periods of two previous candidates to HMXBs and the discovery of
two new candidates (Sasaki et al.\ 2003). The wide field allows a large
number of sources to be observed at the same time, considerably
improving our knowledge of HMXB populations.

\begin{figure}
\label{Haberl}
\caption{ Top: False colour image of a field in the LMC generated from {\sc
epic}-pn data (red, green and blue represent images in the 0.3-1.0 keV,
1.0-2.0 keV and 2.0-7.5 keV bands respectively).
 Bottom: Grey-scale image of the same field showing the whole 0.3-7.5 keV
band and the identification of several sources (from Haberl et al. 2003). }  
\epsfig{file=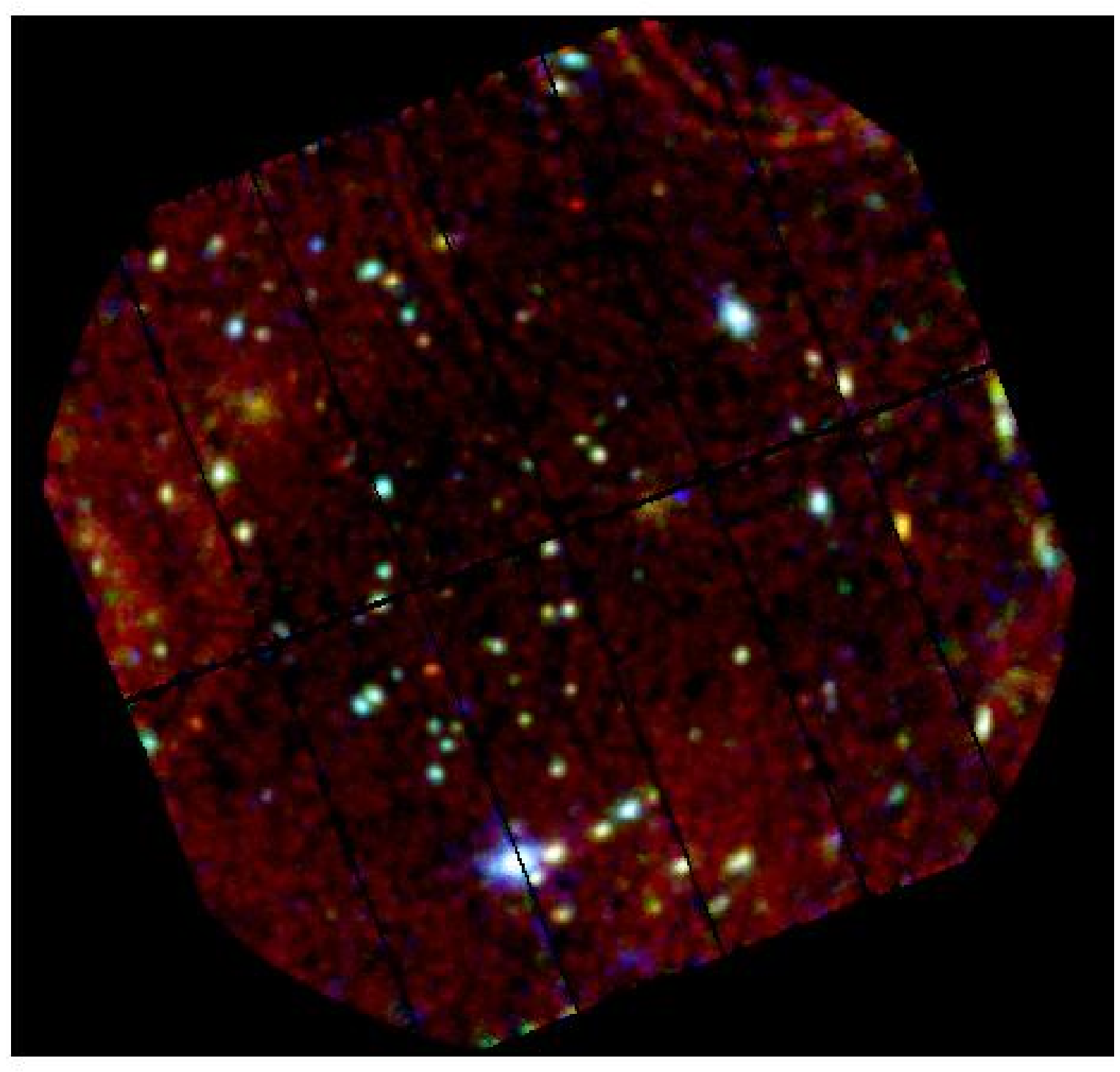,width=8cm,angle=0}
\epsfig{file=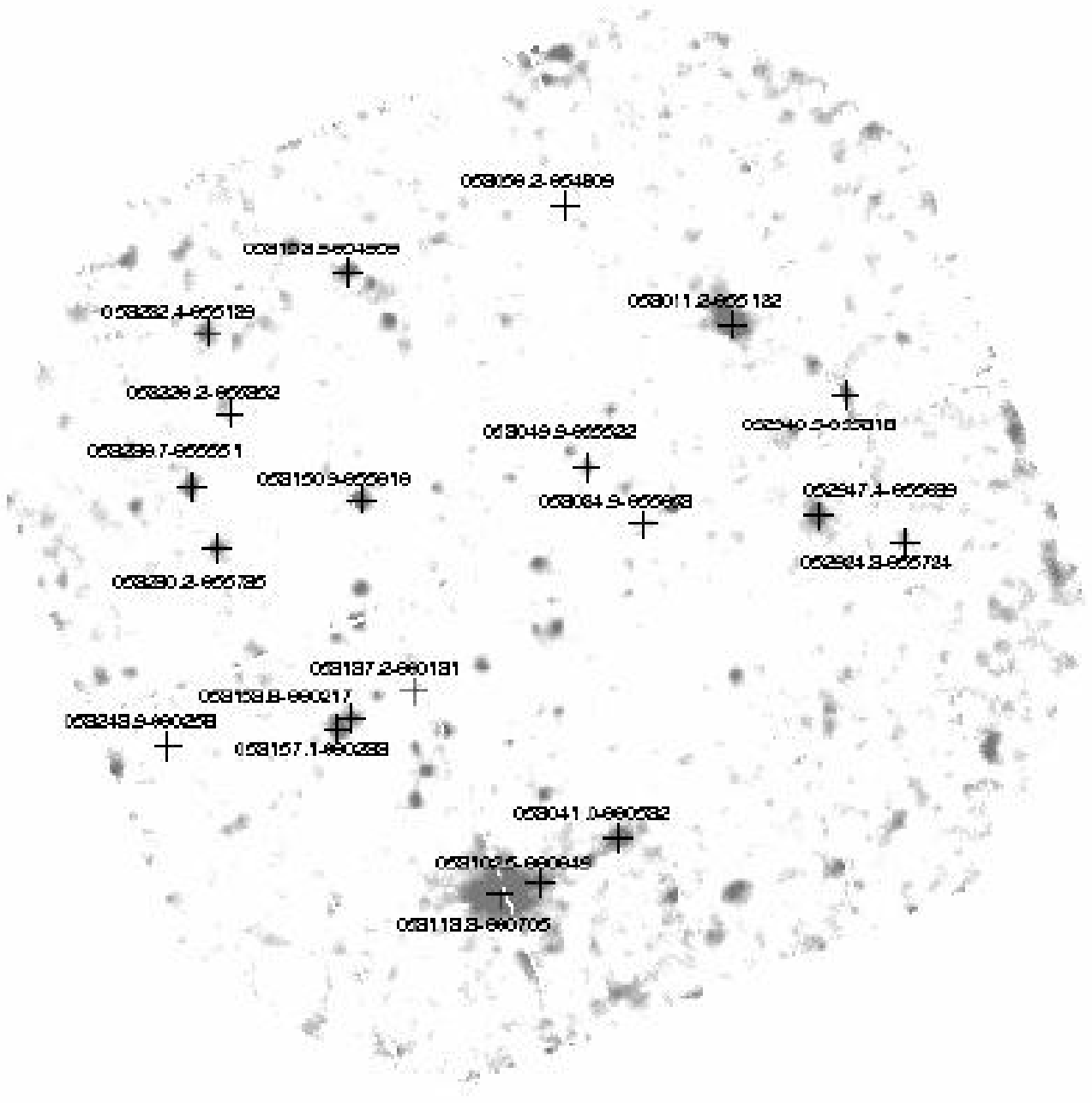, width=7cm,angle=0}
\end{figure}

{\it XMM-Newton} can also extend this sort of work to more distant
galaxies. Pointings to M 31 have allowed the detection of sources down
to a luminosity of $L_{{\rm x}} = 6\times10^{35}\:$erg s$^{-1}$
(Osborne et al.\ 2001), resulting in the characterisation of the whole
XRB population. As expected, the bulge population is dominated by
bright LMXBs, while the disk population appears to be composed mostly
of young HMXBs (Trudolyubov et al.\ 2002a). The temporal evolution of
some bright transients has been studied with both {\it XMM-Newton} and
{\it Chandra}  (Trudolyubov et al.\ 2001). Among them, a bright LMXB has
been
detected in a globular cluster of the M 31 system (Trudolyubov et
al.\ 2002b). All this work suggests an
overall galactic XRB population similar to that of the Milky Way.

Even further, {\it XMM-Newton} has identified the first eclipsing
X-ray binary outside the Local Group (Pietsch et al.\ 2003). This is
RX J004717.4$-$251811 in the nearby starburst galaxy NGC 253.

\subsection{Ultraluminous X-ray sources}
One of the most interesting problems in extragalactic X-ray
Astrophysics is the existence of Ultraluminous X-ray sources
(ULXs). These sources are variable on short timescales, but their
luminosities are typically much higher than those observed in Galactic
and Magellanic Cloud XRBs, approaching in many cases $L_{{\rm
x}}\approx10^{40}\:$erg s$^{-1}$. As this luminosity is much higher
than the Eddington limit for a neutron star or a low-mass black hole,
it seems to require massive black holes with masses of the order of
$\sim 100M_{\odot}$ (generally called intermediate-mass black holes,
as their masses are between those of stellar-mass black holes and the
supermassive objects at the heart of galaxies). Severe problems are
found when trying to understand the formation mechanism of such black
holes, leading to the thought that perhaps ULXs can be explained as
normal XRBs in which the X-ray emission is beamed towards the
observer, requiring thus much lower fluxes (King et al.\ 2001).

A very important result obtained with {\it XMM-Newton} has been the
discovery of quasi-periodic modulation in the flux from an ULX in M82
(Strohmayer \& Mushotzky 2003). Such modulation must arise in the
immediate vicinity of the compact object and represents evidence
against beaming. Unfortunately, because of the relatively low spatial
resolution, the identification of the ULX is not completely certain
and a simultaneous observation with {\it Chandra} seems desirable.

Further evidence supporting intermediate-mass black holes has been
collected from analysis of the spectra of two ULXs in the nearby
spiral galaxy NGC 1313. The blackbody temperatures inferred from the
fits are a factor of ten lower than those of typical SXTs, suggesting
rather more massive black holes (Miller et al.\ 2003c)

\section{Clusters of galaxies}

One of the very first discoveries of X-ray astronomy was that clusters
of galaxies are strong X-ray emitters (see Sarazin 1986 for an early
review and Mushotzky 2001 for a much more up-to-date compilation).  Their
X-ray spectrum is well fitted by plasma emission at a temperature of
$10^7-10^8\, {\rm K}$, which includes thermal bremsstrahlung and line
emission, most notably the Fe K emission line at $\sim 6.7\, {\rm
keV}$.  The inferred Fe abundance is about $\sim 0.3-0.5$ solar, with
possible gradients across the cluster, but remarkable homogeneity
across the cluster population. Clusters (and groups) of galaxies are
therefore filled with enriched gas (likely deposited by the mass loss
of the member galaxies), which appears to be trapped in the cluster
potential well.

The intracluster medium appears to be close to hydrostatic equilibrium
Except for the core (with a few $\times 100\, {\rm kpc}$), cluster gas
is isothermal (or perhaps with a slowly decaying temperature) out to
the distances where X-ray emission can be detected (less than the
virial radius).  Relaxed clusters often exhibit a {\it cooling flow}
phenomenon, whereby the gas in the central part of the cluster is
significantly cooler and the density higher probably due to a highly
subsonic inflow amounting typically to $\sim 100\, {\rm M}_{\odot}\,
{\rm yr}^{-1}$ (see Fabian 1994 for an extensive review).

\subsection{The physical structure of the intracluster medium}

One of the best studied clusters with {\it XMM-Newton} is A1795 ($z=0.063$).
This cluster shows a smooth gas density profile along the lines of the
$\beta$ profile adopted for many clusters (Arnaud et al. 2001). The
temperature of the gas is seen to be constant from 0.1 to 0.4 virial
radii, but dropping significantly at smaller distances from the
cluster centre. This is now seen as a common feature in many clusters,
like A496 (Tamura et al. 2001), A1413 (Pratt \& Arnaud 2002).

The possibility of mapping the intracluster gas structure with
unprecedent detail is now also opening big questions on its physical
state.  In their detailed analysis of the $z=0.143$ A1413 cluster,
Pratt \& Arnaud (2002) argue that the gas departs from hydrostatic
equilibrium when approaching the virial radius, as expected from the
very long timescales involved.  They also find evidence for a cuspy
central density profile (at variance of the usually assumed
$\beta$-profile). The mass profile of the cluster, as derived under
the assumption of hydrostatic equilibrium, is not far from the
predictions of standard Cold Dark Matter computations.

Another important contribution of {\it XMM-Newton} to cluster science has
been illustrated with the observations of the $z=0.54$ cluster
CL0016+16, a working horse for the use of the Sunyaev-Zel'dovich effect
to determine accurate values of the Hubble constant.  Worrall \&
Birkinshaw (2003) have been able to derive accurate values for the gas
temperature (to within 2.5\%) and of the emission measure (electron
density integrated along the line of sight) to within a similar
accuracy.  The subsequent revised value of $H_0=68\pm 8\, {\rm km}\,
{\rm s}^{-1}$ is now in agreement with the latest determinations using
the {\it Hubble Space Telescope} and the Cosmic Microwave Background
power spectrum obtained by WMAP.

\subsection{Building up clusters}

For many years it has been thought that clusters build up from smaller
structures, as predicted by popular cosmological scenarios.  For
instance, evidence for hierarchical merging in the Coma cluster was
provided by {\it ROSAT} observations of X-ray emission of a merging
sub-structure (White, Briel \& Henry 1993).

With its much improved sensitivity, {\it XMM-Newton} is able to map
intracluster gas down to very low surface brightness limits and
therefore to strengthen or disprove the merging hypothesis.  Indeed,
the very first EPIC images of the Coma cluster (Briel et al. 2001)
confirm the presence of the probably infalling lump from the SW
(around NGC 4839) already detected by {\it ROSAT}, but they also find
a further lump probably ahead of  this one on its infall into the Coma
cluster. N-body simulations do predict indeed that the infall into
large clusters of galaxies proceeds along filamentary structures!  The
structure of the gas around NGC 4839 does show clear traces of
being into its first infall into Coma (Neumann et al. 2001).  The tail
of the X-ray emission is very hot ($\sim 4.5\, {\rm keV}$) and indeed
hotter than the temperature of the galaxy itself. Together with a
displacement of the X-ray emission with respect to the galaxy NGC 4839
(well placed at the centre of the optical group) this provides
confirmation of the subgroup gas being ram pressure stripped in its
infall into the Coma cluster.

Evidence for filamentary structures infalling onto clusters is also
seen in {\it XMM-Newton} observations of the cluster A 85 (Durret et
al 2003). Early {\it ROSAT} reports on a large $\sim 4\, {\rm Mpc}$
filamentary structure towards the south of the central cD galaxy of A
85 are confirmed, together with a determination of the gas temperature
of around $\sim 2\, {\rm keV}$. This suggests that the filamentary
structure is made of a chain of several galaxy groups.

Further evidence in favour of clusters being built by merging blocks
comes from observations of distant clusters. De Filippis et al. (2003)
observed the cluster CL 0939+4713 at $z=0.47$ and showed that it
exhibits significant gas structure, with the gas in between the two
dominant galaxies being the hottest.  This implies a largely
non-relaxed state with a likely collision between the two merging
blocks within a few million years.  Hashimoto et al. (2002) observed
the very distant cluster RX J1053.7+5735 at $z=1.26$ and also detected
a double structure suggestive of a merger.

Therefore, {\it XMM-Newton} has brought evidence for merging activity
in clusters of galaxies, particularly at significant redshifts.
Assessing the incidence of mergers in the building up of galaxy
clusters (there are examples of dynamically relaxed clusters at high
redshift, see Arnaud et al. 2002, for observations of RX J1120.1+4318
at $z=0.6$) has just started but will ultimately provide the most
direct evidence on how large-scale structure forms in the Universe.

\subsection{Challenging cooling flows}

Perhaps one of the most unexpected discoveries by {\it XMM-Newton} has been
the challenge to the standard picture of the cooling flow phenomenon.
The cooling time in the centres of galaxy clusters is significantly
smaller than the age of the Universe, and therefore that gas will cool
if no other phenomena prevent it (see, e.g., Fabian 1994 for a
comprehensive review). Maintaining hydrostatic equilibrium with the
hotter outer gas implies that gas should be steadily flowing inwards,
increasing the density towards the cluster centre. Mass deposition
rates are of the order of $\sim 100-1000\, {\rm M}_{\odot}\, {\rm
yr}^{-1}$, which integrated over a Hubble time result in a substantial
contribution to the mass of a large galaxy.  Cooling flows manifest
themselves as a peaked surface emissivity at the centre of the
cluster, where the X-ray spectrum shows cooler gas.  They occur almost
invariably in cD clusters (i.e., those dominated by a large
central-dominant early type galaxy), but very rarely happen in more
non-relaxed structures, possibly due to the disruption of cooling
flows by mergers.

\begin{figure}
\label{M87}
\caption{{\it XMM-Newton} RGS spectrum of the core of the Virgo
  cluster around the M87 galaxy (from Sakelliou et al. 2002).  The
  discrepancy between the standard cooling flow model prediction (top
  green line) and the data (bottom lines) is most evident by the lack
  of Fe L lines around 12-17 \AA.}
  \epsfig{file=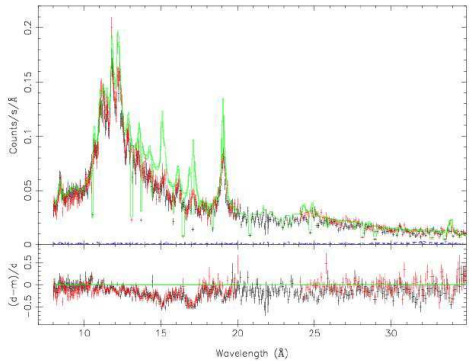,width=6cm,angle=270}
\end{figure}

Previous to the operation of {\it Chandra} and {\it XMM-Newton} there
were doubts about the physics governing cooling flows.  Conduction need
to be suppressed to prevent the gas from thermalising in the hotter
cluster environment, but lower temperature gas was unambiguously seen.  The
most intriguing question was (and still is) where does all this gas go.
cD galaxies present at the centres of cooling flow clusters are red
and, although they do form some stars (see the extensive work by Cardiel
1999 on star formation in cooling flow clusters), they certainly do not
form $\sim 100-1000\, {\rm M}_{\odot}$ {\it normal} stars every year.
A high pressure ambient as that present in cooling flows could favour
the formation of low-mass stars.

A major {\it XMM-Newton} result has been to show that the gas in
cooling flow clusters does not cool below a temperature which is 1/2
to 1/3 of the outer intracluster medium temperature (see, e.g.,
Kaastra et al. 2001, Peterson et al. 2001, Tamura et al. 2001, Peterson
et al. 2003, Sakelliou et al. 2002).  Cluster gas at temperatures $\sim
1$~keV should cool mostly through Fe L lines, which should be very
strong.  Instead they are almost absent of the {\it XMM-Newton} RGS
spectra. The amount of gas that cools is probably 1/10 of the gas
expected from simple cooling flow models (Fabian \& Allen 2003),
suggesting a much modest role of cooling flows in galaxy formation
(see also Molendi \& Pizzolatto 2001).

The reasons for that are still unclear.  Both conduction and heating
by the central source (usually a powerful AGN) have been revived.
Conduction should probably be decreased by the presence of magnetic
fields, as a temperature gradient is indeed seen.   Heating is also
non-trivial, as it should deposit heat uniformly over a large volume,
but on the contrary {\it Chandra} images show lots of structure in
cluster cores.  

A further complicating point is that, despite the fact that the amount
of gas seen at temperatures of millions of degrees is much smaller
than predicted by the simplest cooling flow models, {\it FUSE}
observations reveal in some cooling flow clusters large OVI emission
lines (Oegerle et al. 2001, Bregman 2003, priv comm). The amount of gas
cooling at a few $10^5\, $K is paradoxically in agreement with the
{\it cooling flow} models, but the few million degree gas is
missing. There is clearly much to learn in the coming years from this
phenomenon.

\section{Active Galactic Nuclei}

Active Galactic Nuclei (AGN) are the most populous class of X-ray
sources in the Universe, particularly at high galactic latitude. The
large variety of AGN manifestations results in distinctive X-ray
emission properties.  Radio-loud AGNs were early claimed to have
flatter X-ray spectra than radio-quiet ones (Wilkes \& Elvis 1987),
with flat spectrum radiosources having an even flatter X-ray spectrum
(Canizares \& White 1989). This is now understood in terms of
radio-loud AGNs being absorbed by cold gas (Sambruna et al. 1999), plus
relativistic beaming affecting the X-ray emission in core-dominated
radio sources.

Radio-quiet type 1 AGN (Seyfert 1 galaxies and QSOs) where optical
broad emission lines are seen and type 2 AGN (Seyfert 2s) where these
broad components are not seen in unpolarized light, should have
different X-ray properties.  Indeed, the simplest version of the AGN
unified scheme (Antonucci 1993), where the type 1/type 2 dichotomy
results from different viewing directions of the same (or very
similar) central engine, predicts that X-rays emerging from the
nucleus would be directly seen in type 1 AGNs but absorbed/scattered
in type 2 AGNs.

The standard AGN model, i.e., that of a supermassive black hole fed by
a geometrically thin accretion disk and collimated jets along the
rotation axis, has found some of its best support in the X-ray
observations.  Hard X-ray photons at a rate of $10^{44}\, {\rm erg}\,
{\rm s}^{-1}$ are very difficult to produce by other means in a
galactic centre.  Additionally, some spectral features, as a broad
relativistic Fe K$\alpha$ emission discussed below, are distinctive
tracers of accretion onto massive compact objects.

X-ray emission from AGN is characterised by an underlying power law
with photon index $\Gamma\sim 1.5-2$ (on the steep side for
radio-quiet AGN), which probably rolls over at energies $\sim 100\,
{\rm keV}$ (see the review by Mushotzky et al. 1993
summarizing the situation pre-{\it XMM-Newton}). For radio-quiet
objects and those which are radio-loud but the beaming towards the
observer is not important, this power law is interpreted as the
Compton up-scattering of the UV photons produced in the accretion disk
by the disk's relativistic electron atmosphere.

An Fe K$\alpha$ emission line at 6.4-6.7 keV is the most prominent
emission feature, along with a high-energy `bump' at $>20\, {\rm keV}$
(Pounds et al. 1990), both of them probably arising from reflection of
the X-rays in some thick material. For Seyfert 1 galaxies the
equivalent width of the Fe line lies in the range 100-300 eV, while
for Seyfert 2 galaxies the emission line is much stronger (up to 1 keV
of equivalent width), in particular in Compton-thick Seyfert 2s. The
most striking feature of the Fe line is that it is very broad
(inferred Doppler velocities up to $\sim c/3$). In the best studied
case of the Seyfert 1 galaxy MCG-6-30-15 it exhibits a
relativistically broadened shape interpreted in terms of Doppler
shifts produced by the disk rotation, aberration and gravitational
redshift in the red horn due to the proximity to the black hole's
event horizon (Tanaka et al. 1995).  Similar shapes were observed in
other Seyfert 1 galaxies, suggesting that this is a relatively general
feature of type 1 AGNs.  In the best studied Seyfert 2 galaxies
(e.g. NGC 1068), {\it ASCA} showed a more complex emission line
structure (Iwasawa et al. 1997).

AGN often exhibit photoelectric absorption features in excess of those
produced by cold gas in our Galaxy.  Radio-loud AGNs at high
redshifts, often exhibit large cold (neutral) absorbing gas columns
(Cappi et al. 1997) but in radio-quiet AGN the
situation is more complex.  In this case, type 1 AGN appear to have
little neutral absorbing gas (Nandra \& Pounds 1994), but they often
exhibit signs of ionised gas along the line of sight in the form of
absorption edges, most notably OVII K at 0.74 keV and OVIII K at 0.87
keV (Reynolds 1997, George et al. 1998).  On the contrary Seyfert 2 AGN
usually display large absorbing columns of neutral gas (Smith \& Done
1996), which is attributed to the ``dusty torus'' that the unified AGN
scheme predicts to hide the broad line region. The average good
correlation between optical spectral properties of radio quiet AGN and
the amount of neutral absorbing gas along the line of sight has been
clearly shown in Risaliti et al. (1999) on the basis of an
[OIII]-selected sample of AGN (the [OIII] emissivity is believed to
originate further away than any obscuring material in AGN, and
therefore is regarded as an isotropic measure of the true intrinsic
AGN luminosity).

\subsection{The complex Fe line}

{\it Chandra} and to a larger extent {\it XMM-Newton} have provided a
qualitative leap forward in the study of Fe K lines, due to the large
increase (in particular for {\it XMM-Newton}) of effective area at
$\sim 6\, {\rm keV}$ with respect to previous missions while
preserving or improving the spectral resolution.

The best-studied Seyfert 1 galaxy MCG--6-30-15 has indeed been the
subject of several long {\it XMM-Newton} observations.  Fabian et al.
(2002) discuss a 300 ks observation, where they confirm the existence
of a very broad Fe line feature.  A careful physical modeling of this
feature shows that the line emissivity comes in two parts: the blue
part of the line arises from the outer accretion disk (radius $> 6
r_g$, where $r_g=GM/c^2$ is the gravitational radius of the black hole
with mass $M$), but to account for the very extended red tail
displayed by the line (extending down to 4 keV) a very concentrated
emission at much lower radii is required.  This unambiguously confirms
early claims based on {\it ASCA}  that reflection features arise from
radii as close as $\sim 2\, r_g$ from the black hole, requiring a
rapidly rotating (Kerr) hole. In fact Wilms (2001) caught the same
source in the ``low'' state where the red wing of the Fe line
dominates, suggesting extraction of energy from the spinning
black hole through magnetic fields.  In any case it is
clear that the Fe line diagnostic of MCG--6-30-15 can only be
understood in the context of reflection in an accretion disk around a
Kerr black hole.

Observations of many other Seyfert galaxies have shown that the
profile of the Fe line varies very much from source to source.  In
general, the reflection Fe line arises either in the inner part of the
accretion disk, where it is likely to be highly ionised and broad,
and/or at much more distant regions (perhaps the ``dusty torus'')
where it betrays mostly neutral Fe with a narrow line. Prototypical
cases showing both a narrow component at 6.4 keV and a broad component
centered at 6.7 keV are Mkn 205 (Reeves et al. 2001) and NGC 5506 (Matt
et al. 2001).

\begin{figure}
\label{MCG}
\caption{{\it XMM-Newton} EPIC-MOS spectrum of the Seyfert 1 galaxy
  MCG-6-30-15, around the Fe emission line, fitted to a relativistic
  disk model (from Fabian et al. 2002).}  
\epsfig{file=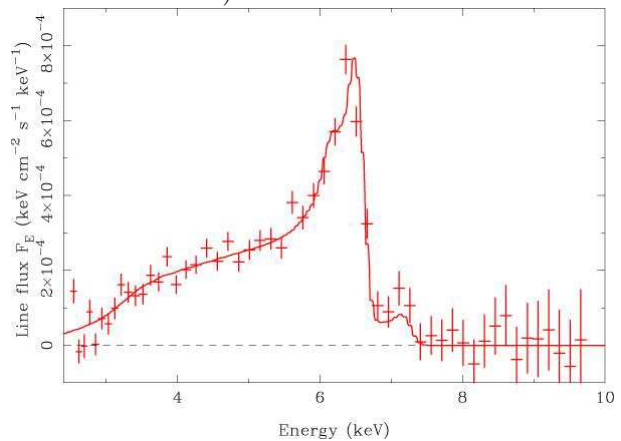,width=8cm,angle=0}
\end{figure}

Interestingly, one of the cases claimed to have a broad Fe line with
{\it ASCA} data, NGC 5548, does only show a neutral narrow line in
{\it XMM-Newton} observations (Pounds et al. 2003). Although the Fe
line appears to vary less than the continuum in some of the
best-studied Seyferts (such as MCG--6-30-15), there are reports of
varying Fe lines.  Guainazzi (2003) studies non-simultaneous {\it
ASCA}, {\it XMM-Newton} and {\it BeppoSAX} observations of
ESO198-G24, and he reports a broad component appearing and then fading
away along with a narrow component.

It must also be stressed that the ``holy grail'' of using
reverberation mapping (i.e., the response of reflected components,
such as the Fe line, to continuum variations) to measure
the mass of the black holes involved in AGN, has turned out to be
rather complex.  The Fe line intensity does not trivially follow the
continuum variations and therefore much work is needed before this
technique can be applied.

\subsection{Soft X-ray broad lines?}

A very lively debate has arisen since the launch of {\it XMM-Newton}
on whether or not there are weak, relativistically broadened emission
lines corresponding to C, N and O at soft X-ray energies.  This stems
from the use of the grating spectrographs both in {\it XMM-Newton} and
{\it Chandra}, which provide spectral resolutions of the order of
200-500.  Branduardi-Raymont et al. (2001) were the first to suggest
the presence of these features in the RGS spectra of MCG-6-30-15 and
Mkn 766, as an alternative to a standard warm absorber picture.  Using
a {\it Chandra} observation of MCG-6-30-15, Lee et al. (2001)
identified a number of spectral features that were expected from a
warm absorber, and even some of them only expected in Fe atoms trapped
in dust grains, supporting the dusty warm absorber picture and no
evidence for broad emission lines. Mason et al. (2003) have recently
reported an analysis of RGS data on Mkn 766, where it is shown that
broad relativistic emission lines of C, N and O give a better fit to
the data than a dusty warm absorber.  As of today, higher signal to
noise {\it XMM-Newton} data over a larger bandpass  appear to favour the
presence of relativistic soft X-ray emission lines, and higher
spectral resolution {\it Chandra} data appear to favour the dusty warm
absorber interpretation. But the story continues $\ldots$

\subsection{The complexity of the ionized absorbers}

The higher spectral resolution provided by RGS has enabled the
possibility of studying the ionized absorbers in AGN with much more
detail.  Sako et al. (2001) conducted the first of such detailed
analyses on the QSO IRAS 13349+2438, which was already suspected from
the combined {\it ROSAT}, {\it ASCA} and optical data to have a
complex absorption structure along the line of sight.  The RGS
spectrum shows a large number of absorption lines of C,N,O,Ne with one and
two electrons, as well as a number of Fe lines in various ionisation
states. Most remarkably, an Unresolved Transition Array (UTA) of
inner-shell lines of Fe M is detected for the first time at X-ray
wavelengths, which might be confused, at the lower resolution of
non-dispersive CCD imaging spectrographs, with an absorption
edge. Absorption lines of Fe are detected in Fe VII - Fe XII and in Fe
XVII - Fe XX, but not in the intermediate ionisation states Fe XIII -
Fe XVI. This suggests the existence of two distinct absorbers, a low
ionisation one, which happens to be outflowing with a velocity $\sim
400\, {\rm km}\, {\rm s}^{-1}$ and a high ionisation absorber
basically at rest with the QSO.  The low ionization outflowing
absorber implies (for normal dust to gas ratio) a reddening coincident
with the optical one, but the high ionisation absorber, with a much
larger column density, does not appear to have any dust.

\begin{figure}
\label{IRAS}
\caption{{\it XMM-Newton} RGS spectrum of the QSO IRAS 13349+2438,
  showing the complexity of the ionized absorber and the presence of a
  Unresolved Transition Array around 16-17 \AA.}  
\epsfig{file=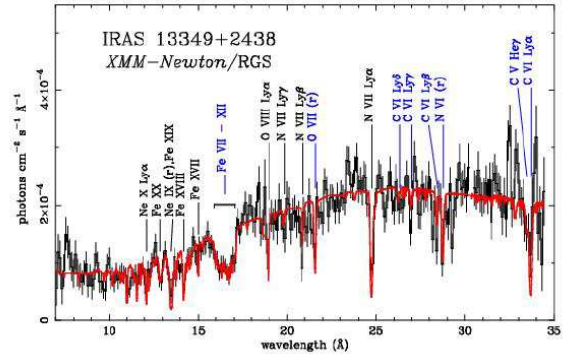,width=8cm,angle=0}
\end{figure}

Similar deep studies have been conducted on other targets (e.g.,
Blustin et al. 2002), revealing almost invariably the presence of UTAs,
and a complex (at least two-phase) ionisation structure for the
absorbers.  It must be stressed that we are likely probing the gas in
between the broad-line region and the ``obscuring torus'' in a manner
that no other observational strategy could do.

\section{Surveys and the X-ray background}

The existence of a cosmic X-ray background (CXRB) was among the very
first discoveries of X-ray Astronomy (Giacconi et al. 1962).  Its
spectrum was well measured by the HEAO-1 mission (Marshall et al. 1980,
Gruber 1992) and fits remarkably well a thermal bremsstrahlung
model at a temperature of $\sim 30\, {\rm keV}$. At galactic latitudes
$>20^{\circ}$ the CXRB is remarkably isotropic claiming for a true
cosmological origin (see Fabian \& Barcons 1992 for a review). Most of
the energy density in the CXRB resides at $\sim 30\, {\rm keV}$, an
energy which is not accessible to sensitive enough instrumentation in
the current era.  Below 1 keV the Galaxy (and probably a local hot
bubble) dominates the X-ray intensity, and the extragalactic component
is shielded by the interstellar medium, making it difficult to
measure. In spite of this, deep {\it ROSAT} observations
resolved $\sim 70\%$ of the CXRB in the 1-2 keV (Hasinger et al. 1998)
while {\it Chandra} has resolved practically 90\% of this and $\sim
70\%$ of the 2-7 keV CXRB (Mushotzky et al. 2000).  

The lack of spectral distortions of the Cosmic Microwave
Background ruled out the presence of substantial amounts of
intergalactic gas at the very high temperatures needed to produce the
CXRB through thermal bremsstrahlung (Barcons et al.
1991). Finding the sources that produce the CXRB, their astrophysical
nature, their cosmic evolution, spatial clustering and so on has been
since then the main objective of CXRB studies. 

The most popular models for the CXRB make use of the unified model for
AGN, and assume a mixture of unabsorbed (presumably type 1) and
absorbed (presumably type 2) objects as a function of redshift $z$
(Comastri et al. 1995, Gilli et al. 2001).  

Medium sensitivity (Mason et al. 2000) and deep (Lehmann et al. 2002)
surveys carried out with {\it ROSAT} revealed that at soft X-ray
energies the X-ray sky is dominated by AGNs, most of which are type 1
Seyferts and QSOs.  A small fraction of the AGN were type 2 and other
narrow emission line galaxies (e.g., starburst galaxies). However, due
to its bandpass limited to soft X-ray photons {\it ROSAT} missed most
of the sources absorbed by H columns in excess to $10^{21}\, {\rm
cm}^{-2}$. 

The large field of view of the EPIC cameras ($0.5^{\circ}$ in
diameter) makes {\it XMM-Newton} a very powerful tool to conduct X-ray
surveys.  The Survey Science Centre (SSC) is carrying out a series of
serendipitous surveys in order to characterise the content of the
X-ray sky at different depths.  At high Galactic latitudes, 3 samples
of $\sim 1000$ sources each, selected at 0.5-4.5~keV fluxes of $\sim
10^{-13}$, $\sim 10^{-14}$ and $\sim 10^{-15}\, {\rm erg}\, {\rm
cm}^{-2}\, {\rm s}^{-1}$ will provide a major handle on this task. A
number of surveys are being conducted besides the SSC ones, most of
which are discussed in the proceedings of the workshop {\it X-ray
Surveys, in the light of the new observatories} held in Santander
during 4-6 September 2002 (Astronomische Nachrichten, vol 324, issues
1 and 2).

The AXIS programme\footnote{AXIS is an International Time Programme at
the Observatorio del Roque de Los Muchachos in La Palma, which was
granted 85 observing nights spread on the INT, NOT, TNG and WHT
telescopes from April 2000 through April 2002.  See
http://venus.ifca.unican.es/\~{} xray/AXIS} is providing a substantial
leap forward in the {\it XMM-Newton} SSC medium sensitivity
survey. This is a flux limited survey at 
 $2\times 10^{-14}\, {\rm erg}\, {\rm cm}^{-2}\, {\rm
s}^{-1}$, in the 0.5-4.5 keV band, with a source density of $\sim 100\,
\deg^{-2}$, which has the goal of covering $\sim 5\, \deg^{-2}$ both
in the North and in the South. Preliminary results were presented in Barcons et al.
(2002) and an update can be found in Barcons et al. (2003a).  

\begin{figure}
\label{LH}
\caption{{\it XMM-Newton} EPIC-pn image of the Lockman Hole, totalling
an exposure time approaching 1 Ms (about one week).}  
\epsfig{file=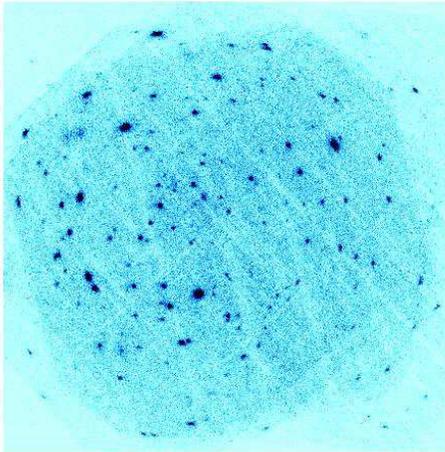,width=8cm,angle=0}
\end{figure}

The deepest X-ray survey conducted by {\it XMM-Newton} so far is in the
Lockman Hole area, with a total exposure approaching 1000 ks (adding
the payload verification, AO-1 and AO-2 data).  The
scope of this survey is not only to detect the faintest sources,
particularly at hard X-ray energies where confusion is still
unimportant and {\it XMM-Newton} is 5-10 times more efficient than
{\it Chandra}, but more importantly to obtain X-ray spectra of the
faintest sources. A preliminary account, based on 100 ks data, is
presented in Hasinger et al. (2001).

\subsection{Challenging the AGN unified scheme}

{\it XMM-Newton} is also providing evidence that the one to one
association of type 1 AGNs with unabsorbed sources and type 2 AGNs
with absorbed sources fails clearly in some cases. For example Barcons
et al. (2003b) have shown that the H1320+551 z=0.068 Seyfert 1.9
galaxy (and therefore with a broad line region putatively obscured)
does not show any signs of neutral  or ionised  gas along the line of
sight. This is in direct contradiction to the unified AGN model.

On the opposite side, type 1 QSOs are found which display large
absorbing columns of neutral gas (e.g., Georgantopoulos et al. 2003),
while none or very little was expected.  A possible explanation to
this is that X-ray absorption occurs very close to the central engine
where dust just sublimates and therefore the optical spectrum is
unaffected by the surrounding material. This might be a non-isolated
situation, as explained by Mateos et al. (2003) in an analysis of the
X-ray spectra of XMS sources, where a fraction of type 1 AGNs do
require absorbing columns along the line of sight.

\subsection{Faint and absorbed sources}

If anything, the deep observation of Hasinger et al. (2001)
demonstrates that the basic picture for the origin of the CXRB, based
on absorbed sources, is roughly correct.  Indeed, some $\sim 40\%$ of
the faint sources detected show some hint of absorption in their X-ray
colours, in agreement with the general expectations. Quantitatively,
however, the situation is still far from clear.

One of the most striking results obtained from the optical
identification of {\it Chandra} deep surveys is that the redshift
distribution of type 2 AGN is peaked at significantly lower redshifts
($z\sim 0.8$) than that of type 1 AGN ($z\sim 1.7$) (Hasinger 
2002). Similarly, in the XMS all type 2 AGNs are concentrated at low
redshifts, while the type 1s reach $z\sim 3$. 

This probably means that the models for CXRB need some
revision. Following this, Franceschini, Braito \& Fadda (2002) have
suggested that type 1 AGN arise in the rare highest peaks of the
density field in the Universe, but that type 2 AGN result from the
evolution and merging of the very massive galaxies, a process that
occurs later in the cosmic history.

\subsection{Optically dull X-ray bright galaxies and obscured accretion}

A still intriguing issue raised by the new observatories and studied
in detail by {\it XMM-Newton} is the existence of optically inactive
but X-ray luminous galaxies.  Such objects were first highlighted in a
{\it Chandra} deep survey by Mushotzky et al. (2000), as early type
galaxies with a large X-ray to optical ratio.  Comastri et al. (2002)
conducted a multi-wavelength study of one of these objects discovered
in their HELLAS2XMM survey, which shows no optical or infrared
emission lines, but does show a rather hard X-ray spectrum.  They
concluded that the counterpart galaxy at $z=0.159$ hosts a completely
obscured active nucleus.

Severgnini et al. (2003) have recently reported on the study of 3 of
these objects discovered in the SSC bright source survey.  Their X-ray
properties are varied, but in all cases reflecting clearly the
presence of an X-ray luminous AGN.  In that work it is shown that the
emission lines related to the more or less obscured AGN are diluted
with the galaxy light and therefore very weak and inconspicuous in some
cases.  

All this appears to support previous suspicions that most, if not all,
galaxies host a black hole in their center, many of them active but
obscured.  Obscured accretion is probably 3 times more frequent in the
Universe than unobscured accretion, if the CXRB models are roughly
correct.  A detailed accounting on the black hole density can then be
inferred by assuming a radiative efficiency of the black holes
(typically 10\%) and compared to other independent estimates of the
local black hole density.  Fabian (2003) concludes that the estimate
based on the CXRB and with what is known from the redshift distribution
of obscured sources, converges towards $\sim 4\times 10^{5}\, {\rm
M}_{\odot}\, {\rm Mpc}^{-3}$, a value which is consistent with local
estimates. Obscured accretion onto black holes is therefore
important, although the ``hidden'' black hole mass is of the same order
than that seen in unobscured AGN.

\section{Gamma-ray burst afterglows}

{\it XMM-Newton} is making a big effort in observing Gamma Ray Bursts
(GRB) following alerts provided by other higher-energy observatories.
The absolute minimum reaction time reached is 6-8 hours after the
alert.  In all observations (conducted as Target Of Opportunity
observations, and therefore immediately public) the GRB afterglow has
been detected and much has been learnt on this phenomenon from the
X-ray data.

Perhaps the most interesting case is that of GRB011211 (identified to
occur in a galaxy of $\sim 25^{\rm mag}$ at $z=2.141$) observed by
{\it XMM-Newton} for 27 ks, 11 hours after its detection by {\it
BeppoSAX}.  During the first 10 ks of that observation, the EPIC X-ray
spectrum showed emission lines which Reeves et al. (2002) identified as
Mg XI, Si XIV, SXVI, Ar XVIII and Ca XX but at a significantly
blueshifted velocity $\sim 0.1c$ (the detection confidence of these
lines has been challenged by Rutledge \& Sako 2003 though).  If the
Reeves et al. (2002) interpretation is correct, then this would provide
direct evidence for outflowing material, favouring the model of the
collapse of a massive star for GRBs.

\section*{Acknowledgments} The work reported herein is based partly on
observations obtained with {\it XMM-Newton}, an ESA science mission
with instruments and contributions directly funded by ESA member
states and the USA (NASA).  We are grateful to Fred Jansen, Christian
Motch, Richard Mushotzky and an anonymous referee for comments on an
early version that resulted in substantial improvements. XB
acknowledges partial financial support by the Ministerio de Ciencia y
Tecnolog\'\i a (MCyT), under grant AYA2000-1690.  IN is a researcher of the
programme {\em Ram\'on y Cajal}, funded by the University of Alicante
and the MCyT. He acknowledges partial financial support by the MCyT
under grants ESP2001-4541-PE and ESP2002-04124-C03-03.

\section*{References}

\noindent Antonucci, R.R.J. 1993, Unified models for active galactic
nuclei and quasars, ARAA, 31, 473

\noindent Arnaud, M. et al. 2001, Measuring cluster temperature
profiles with XMM/EPIC, A\&A, 365, L188

\noindent Arnaud, M. et al. 2002, {\it XMM-Newton} observation of the
distant (z=0.6) galaxy cluster RX J1120.1+4318, A\&A, 390, 27

\noindent Audard, M. et al.\ 2003, A study of coronal abundances in RS CVn
binaries, A\&A 398, 1137

\noindent Barcons, X. et al. 1991, The physical state of the
intergalactic medium, Nat, 350, 685

\noindent Barcons, X. et al. 2002, The {\it XMM-Newton} Serendipitous
Survey. II. First results from the AXIS high galactic latitude medium
sensitivity survey, A\&A, 382, 522

\noindent Barcons, X.  et al. 2003a, The {\it XMM-Newton} Survey Science
Centre medium sensitivity survey, AN, 324, 44

\noindent Barcons, X. et al. 2003b, H1320+551: A Seyfert 1.8/1.9
galaxy with an unabsorbed X-ray spectrum, MNRAS, 339, 757

\noindent Becker, W., \& Aschenbach, B. 2003, X-ray observations of neutron
stars and pulsars: First results from {\it XMM-Newton}, in Becker,
W. et al.\, eds., Proceedings of 270th WE-Heraeus seminar on Neutron
stars, pulsars and supernova remnants, MPE Report 278, p. 64
(astro-ph/0208466)

\noindent Becker, W., \& Tr\"umper, J. 1993, Detection of pulsed X-ray
emission from the binary millisecond pulsar J0437$-$4715

\noindent Bhattacharya, D., \& van den Heuvel, E.P.J. 1991, Formation
and evolution of binary and millisecond radio pulsars, Ph.R. 203, 1

\noindent Blustin, A.J. et al. 2002, Multi-wavelength study of the
Seyfert 1 galaxy NGC 3783 with {\it XMM-Newton}, A\&A, 392, 453

\noindent Bocchino, F.,\& Bykov, A.M. 2003, {\it XMM-Newton} study of hard
X-ray sources in IC~443, A\&A 400, 203

\noindent Bocchino, F. et al. 2001, The X-ray nebula of the filled
center supernova remnant 3C~58 and its interaction with the
environment, A\&A 369, 1078

\noindent Branduardi-Raymont, G. et al. 2001, Soft X-ray emission
lines from a relativistic accretion disk in MCG-6-30-15 and Mrk 766,
A\&A, 365, L140

\noindent Briel, U.G. et al. 2001, A mosaic of the Coma cluster of
galaxies with {\it XMM-Newton}, A\&A, 365, L60

\noindent Burwitz, V., et al.\ 2003, The thermal radiation of the
isolated neutron star RX J1856.5$-$375 observed with {\it Chandra} and
{\it XMM-Newton}, A\&A 399, 1109

\noindent Campana, S. 2002, An {\it XMM-Newton} study of the 401 Hz
accreting pulsar SAX J1808.4$-$3658 in quiescence, ApJ 575, L15

\noindent Canizares, C.R. \& White, J.L., 1989, The X-ray spectra of
high-redshift quasars, ApJ, 339, 27

\noindent Cappi, M. et al. 1997, ASCA and {\it ROSAT} X-Ray Spectra of
High-Redshift Radio-loud Quasars, ApJ, 478, 492

\noindent Cardiel, N. 1999, Formaci\'on estelar en galaxias
dominantes de c\'umulos, PhD thesis, Universidad Complutense de Madrid

\noindent Church, M. J., \& Balucinska-Church, M. 2001, Results of a
LMXB survey: Variation in the height of the neutron star blackbody
emission region, A\&A 369, 915

\noindent Cohen, D.H., et al.\ 2003, High-resolution {\it Chandra} spectroscopy
of $\tau$ Scorpii: a narrow-line X-ray spectrum from a hot star

\noindent Comastri, A. et al. 1995, The contribution of AGNs to the
 X-ray background, A\&A, 296, 1

\noindent Comastri, A. et al. 2002, The HELLAS2XMM survey:
II. Multiwavelength observations of P3: an X-ray bright, optically
inactive galaxy, ApJ, 571, 771

\noindent Cottam, J. et al.\ 2001, High-resolution spectroscopy of the
low-mass X-ray binary EXO 0748$-$67, A\&A 365, L277

\noindent Cottam, J., et al.\ 2002,
Gravitationally redshifted absorption lines in the X-ray burst spectra
of a neutron star, Nat. 420, 51

\noindent De Filippis, E. et al. 2003, XMM observation of the
dynamically young galaxy cluster CL 0939+4713, A\&A, in the press
(astro-ph/0304027)

\noindent Durret, F. et al. 2003, An {\it XMM-Newton} view of the extended
"filament" near the cluster of galaxies Abell 85, A\&A, in the press
(astro-ph/0303486)

\noindent Esin, A.E. et al.\ 1998, Spectral transitions in Cygnus X-1
and other black hole X-ray binaries, ApJ 505, 854

\noindent Fabian, A.C. \& Barcons, X. 1992, The origin of the X-ray
background, ARAA, 30, 429

\noindent Fabian, A.C., 1994, Cooling Flows in Clusters of Galaxies,
ARAA, 32, 277

\noindent Fabian, A.C. et al. 2002, A long hard look at MCG-6-30-15
with {\it XMM-Newton}, MNRAS, 335, L1

\noindent Fabian, A.C. 2003, What can be learnt from extragalactic
X-ray surveys?, AN, 324, 4

\noindent Fabian, A.C. \& Allen, S.W. 2003, X-rays from Clusters of
Galaxies, in {\it proceedings of the XXI Texas Symposium on
Relativistic Astrophysics}, in the press (astro-ph/0304020)

\noindent Favata, F. et al.\ 2003, An {\it XMM-Newton}-based X-ray survey of
pre-main sequence stellar emission in the L1551 star-forming complex,
A\&A 403, 187

\noindent Feigelson, E.D., \& Montmerle, T. 1999, High-energy
processes in young stellar objects, ARA\&A, 37, 363

\noindent Feldmeier, A., et al.\ 1997, The X-ray emission from cooling
zones in O star winds, A\&A 320, 899

\noindent Fleming, T.A., et al.\ 1988, The
relation between X-ray emission and rotation in late-type stars from
the perspective of X-ray selection, ApJ 340, 1011

\noindent Franceschini, A. et al. 2002, Origin of the X-ray
background and AGN unification: new perspectives, MNRAS, 335, L51

\noindent Georgantopoulos, I. et al. 2003, {\it XMM-Newton} observations of
an absorbed z=0.67 QSO: no dusty torus?, MNRAS, in the press
(astro-ph/0302375)

\noindent George, I.M. et al. 1998, ASCA Observations of Seyfert 1
Galaxies. III. The Evidence for Absorption and Emission Due to
Photoionized Gas, ApJS, 114, 73

\noindent Giacconi, R. et al. 1962, Evidence for x Rays From Sources
Outside the Solar System, Phys Rev Lett, 9, 439

\noindent Gilli, R. et al. 2001, Testing current synthesis models of
the X-ray background, A\&A, 366, 407

\noindent Grimm, H.-J., et al.\ 2003, High-mass
X-ray binaries as a star formation rate indicator in distant galaxies,
MNRAS 339, 793

\noindent Gruber, D.E., 1992, The Hard X-Ray Background, in {\it The
X-ray background}, Barcons, X. \& Fabian, A.C. eds, Cambridge
University Press.

\noindent Guainazzi, M., 2003, The history of the iron K-alpha line
profile in the Piccinotti AGN ESO198-G24, A\&A, in the press
(astro-ph/0302117)

\noindent G\"udel, M. et al.\ 2002, Detection of the Neupert effect in
the corona of an RS CVn System with {\it XMM-Newton} and the VLA, ApJ
577, 371

\noindent Haberl, F. et al.\ 2000, A {\it ROSAT} PSPC catalogue of
X-ray sources in the SMC region, A\&AS 142, 41

\noindent Haberl, F., \& Pietsch, W. 2001, The X-ray view of M 33
after {\it ROSAT}, A\&A 373, 438

\noindent Haberl, F., et al.\ 2003, Deep {\it
XMM-Newton} observation of a northern LMC field: I. Selected X-ray
sources, A\&A, in press (astro-ph/0212319)

\noindent Haberl, F. et al.\ 2003, A broad absorption feature in the
X-ray spectrum of the isolated neutron star RBS 1223 (1RXS
J130848.6+212708), A\&A 403, L19

\noindent Hameury, J.-M. et al.\ 2003, {\it XMM-Newton} observations
of two black hole X-ray transients in quiescence, A\&A 399, 631

\noindent Harnden, F.R. et al.\ 1979, Discovery of an X-ray star
association in VI Cyg/Cyg OB2, ApJ 234, L51

\noindent Hashimoto, Y. et al. 2002, {\it XMM-Newton} Observation of a
Distant X-ray Selected Cluster of Galaxies at z=1.26 with Possible
Cluster Interaction, A\&A, 381, 841

\noindent Hasinger, G. et al. 1998, The {\it ROSAT} Deep Survey. I. X-ray
sources in the Lockman Field, A\&A, 329, 482

\noindent Hasinger, G. et al., 2001, {\it XMM-Newton} observation of the
Lockman Hole. I. The X-ray data, A\&A, 365, L45

\noindent Hasinger, G., 2002, The sources of the X-ray background, in
{\it New visions of the Universe in the XMM-Newton and Chandra era},
in the press (astro-ph/0202430)

\noindent Helfand, D. J., \& Moran, E.C. 2001, The hard X-Ray
luminosity of OB star populations: Implications for the contribution
of star formation to the cosmic X-ray background, ApJ, 554, 27

\noindent Hernanz, M., \& Sala, G. 2002, A classical nova, V2487 Oph
1998, seen in X-rays before and after its explosion, Sci 298, 393

\noindent van den Heuvel, E.P.J. et al.\ 1992, Accreting white dwarf
models for CAL 83, CAL 87 and other ultrasoft X-ray sources in the
LMC, A\&A 262, 97

\noindent Howk, J.C. et al.\ 2000, Stagnation and infall of dense
clumps in the stellar wind of $\tau$ Scorpii, ApJ 534, 348

\noindent Iwasawa, K. et al. 1997, The iron K line complex in NGC1068:
implications for X-ray reflection in the nucleus, MNRAS, 289, 443

\noindent Jansen, F. et al. 2001, {\it XMM-Newton} observatory. I. The
spacecraft and operations. A\&A 365, L1

\noindent Jimenez-Garate, M.A. et al.\ 2002, High-resolution X-Ray
spectroscopy of Hercules X-1 with the {\it XMM-Newton} Reflection
Grating Spectrometer: CNO element abundance measurements and density
diagnostics of a photoionized plasma, ApJ 578, 391

\noindent Kaastra, J.S. et al. 2001, {\it XMM-Newton} observations of the
cluster of galaxies Sersic 159-03, A\&A, 365, L99

\noindent Kahn, S.M. et al.\ 2001, High resolution X-ray spectroscopy
of $\zeta$ Puppis with the {\it XMM-Newton} reflection grating
spectrometer, A\&A 365, L312

\noindent King, A.R. et al.\ 2001, Ultraluminous X-Ray sources in
external galaxies, ApJ 552, L109

\noindent King, A.R., et al.\ 2002, The
short-period supersoft source in M31, MNRAS 329, L43

\noindent Koptsevich et al.\ 2003, Deep $BVR$ imaging of the field of
the millisecond pulsar PSR J0030+0451 with the VLT, A\&A 400, 265

\noindent La Palombara, N., \& Mereghetti, S. 2002, Timing analysis of
the core of the Crab-like SNR G21.5$-$0.9, A\&A 383, 916

\noindent Lee, J.C. et al. 2001, Revealing the Dusty Warm Absorber in
  MCG -6-30-15 with the Chandra High-Energy Transmission Grating, ApJ,
  554, L13

\noindent Lewin W.H.G., et al.\ 1993, X-Ray
bursts, SSRv 62, 223

\noindent Lewin W.H.G., et al.\ 1995, X-ray binaries, Cambridge
Astrophysics Series, Cambridge, MA (Cambridge University Press)

\noindent Lorimer, D.R. 2001, Binary and millisecond pulsars at the
new millennium, LRR 4,5 (astro-ph/0104388)

\noindent Marshall, F.E. et al. 1980, The diffuse X-ray background
spectrum from 3 to 50 keV, ApJ, 235, 4

\noindent Mason, K.O. et al. 2000, The {\it ROSAT} International
X-ray/Optical Survey (RIXOS): source catalogue, MNRAS, 311, 456

\noindent Mason, K.O. et al. 2001, The {\it XMM-Newton} optical/UV monitor
telescope, A\&A, 365, L36

\noindent Mason, K.O. et al. 2003,  The X-ray spectrum of the
Seyfert I galaxy Mrk 766: Dusty Warm Absorber or Relativistic Emission
Lines?, ApJ, 582, 95

\noindent Mateos, S. et al. 2003, X-ray spectra of {\it XMM-Newton} AXIS
serendipitous sources, AN, 324, 48

\noindent Matt, G. et al. 2001, The complex iron line of NGC 5506,
A\&A, 377, L31

\noindent Mereghetti, S., et al.\ 2002a, The X-ray
source at the center of the Cassiopeia A supernova remnant, ApJ 569,
275

\noindent Mereghetti, S. et al.\ 2002b, Pulse phase variations of the
X-ray spectral features in the radio-quiet neutron star 1E
1207$-$5209, ApJ 581, 1280

\noindent Mewe, R., et al.\ 2003, High-resolution X-ray spectroscopy
of $\tau$ Scorpii (B0.2V) with {\it XMM-Newton}, A\&A 398, 203

\noindent Miller, J.M. et al.\ 2002a, {\it XMM-Newton} spectroscopy of
the Galactic microquasar GRS 1758$-$258 in the peculiar off/soft
state, ApJ 566, 358

\noindent Miller, J.M. et al.\ 2003b, Evidence of spin and energy
extraction in a Galactic black hole candidate: The {\it
XMM-Newton}/EPIC-pn spectrum of XTE J1650$-$500, ApJ 570, L69

\noindent Miller, J.M. et al.\ 2003c, {\it XMM-Newton} spectroscopy of
the accretion-driven millisecond X-ray pulsar XTE J1751$-$305, ApJ
583, L99

\noindent Miller, J.M. et al.\ 2003, X-ray spectroscopic evidence for
intermediate-mass black holes: Cool accretion disks in two
ultraluminous X-ray sources, ApJ 585, L37

\noindent Molendi, S. \& Pizzolatto, F. 2001, Is the Gas in Cooling
Flows Multiphase?, ApJ, 560, 194

\noindent Motch, C., et al.\ 2002, {\it XMM-Newton}
observations of MR Vel/RX J0925.7$-$4758, A\&A 393, 913

\noindent Mushotzky, R.F. et al. 1993, X-ray spectra and time
variability of active galactic nuclei, ARAA, 31, 717

\noindent Mushotzky, R.F. et al. 2000, Resolving the extragalactic
hard X-ray background, Nat, 404, 459

\noindent Mushotzky, R.F., 2001, Clusters, in {\it The Century of Space
  Science}, Bleeker, J.A.M., Geiss, J. \& Huber, M.C.E. eds, p. 473

\noindent Nandra, K. \& Pounds, K.A. 1994, GINGA Observations of the
X-Ray Spectra of Seyfert Galaxies, MNRAS, 268, 405

\noindent Narayan, R., McClintock, J.E., \& Yi, I. 1996, A new model
for black hole soft X-ray transients in quiescence, ApJ 457, 821

\noindent Neuh\"auser, R. 1997, Low-mass pre-main sequence stars and
their X-ray emissions, Sci 276, 1363

\noindent Neumann, D. et al. 2001, The NGC4839 group falling into the
coma cluster observed by {\it XMM-Newton}, A\&A, 365, L74

\noindent Noyes, R.W. et al.\ 1984, Rotation, convection and magnetic
activity in lower main-sequence stars, ApJ 279, 763

\noindent Oegerle, W.R. et al. 2001, FUSE Observations of
Cooling-Flow Gas in the Galaxy Clusters A1795 and A2597, ApJ, 560, 187

\noindent Oosterbroek, T. et al.\ 2001, {\it BeppoSAX} observation of
the eclipsing dipping X-ray binary X 1658$-$298, A\&A 376, 532

\noindent Osborne, J.P. 2001, The central region of M 31 observed with
{\it XMM-Newton}. II. Variability of the individual sources, A\&A 378,
800

\noindent Oskinova, L.M. et al.\ 2003, The conspicuous absence of
X-ray emission from Carbon-enriched Wolf-Rayet stars, A\&A 402, 755

\noindent \"Ozel, F., \& Psaltis, D. 2003, Spectral lines from
rotating neutron stars, AJ 582, L31

\noindent Paerels, F. et al.\ 2001a, First XMM-Newton observations of
an isolated neutron star: obtained with XMM-Newton, A\&A, 365, L298

\noindent Paerels, F. et al.\ 2001b, A high resolution spectroscopic
observation of CAL 83 with {\it XMM-Newton}/RGS, A\&A 365, L308

\noindent Parmar, A.N. et al.\ 2002, Discovery of narrow X-ray
absorption features from the dipping low-mass X-ray binary X
1624$-$490 with {\it XMM-Newton}, A\&A 386, 910

\noindent Patterson, J. 1984, The evolution of cataclysmic and
low-mass X-ray binaries, ApJS 54, 443

\noindent Patterson, J. 1994, The DQ Herculis stars, PASP 106, 209

\noindent Peterson, J.R. et al. 2001, X-ray imaging-spectroscopy of
Abell 1835, A\&A, 365, L104
 
\noindent Peterson, J.R. et al. 2003, High Resolution Spectroscopy of
14 Cooling-Flow Clusters of Galaxies Using the Reflection Grating
Spectrometers on {\it XMM-Newton}, in {\it New visions of the Universe in
the XMM-Newton and Chandra era}, in the press (astro-ph/0202108)

\noindent Pietsch, W., et al.\ 2003, RX
J004717.4$-$251811: The first eclipsing X-ray binary outside the Local
Group, A\&A 402, 457

\noindent Porquet, D., et al.\ 2003,
{\it XMM-Newton} spectral analysis of the Pulsar Wind Nebula within the
composite SNR G0.9+0.1, A\&A 401, 197

\noindent Pounds, K.A. et al. 1990, X-ray reflection from cold matter
in the nuclei of active galaxies, Nat, 344, 132

\noindent Pounds, K.A. et al. 2003, A simultaneous {\it XMM-Newton} and
{\it BeppoSAX} observation of the archetypal Broad Line Seyfert 1 galaxy NGC
5548, MNRAS, in the press (astro-ph/0210288)

\noindent Pratt, G.W \& Arnaud, M. 2002, The mass profile of A1413
observed with {\it XMM-Newton}: implications for the M-T relation, A\&A,
394, 375

\noindent Ramsay, G. \& Cropper, M.S. 2002, First X-ray observations of the
polar CE Gru, MNRAS 335, 918

\noindent Ramsay, G. \& Cropper, M.S. 2003, {\it XMM-Newton}
observations of the polars EV UMa, RX J1002$-$19 and RX J1007$-$20,
MNRAS 338, 219

\noindent Rasmussen, A.P. et al.\ 2001, The X-ray spectrum of the
supernova remnant 1E 0102.2$-$7219, A\&A 365, L231

\noindent Rauw, G. et al.\ 2003, Phase-resolved X-ray and optical
spectroscopy of the massive binary HD~93403, A\&A 388, 552

\noindent Reeves, J.N. et al. 2001, {\it XMM-Newton} observation of an
unusual iron line in the quasar Markarian 205, A\&A, 365, L134

\noindent Reeves, J.N. et al. 2002, The Signature of Supernova Ejecta
Measured in the X-ray Afterglow of Gamma-Ray Burst 011211, Nat, 416,
512

\noindent Reynolds, C.S. 1997, An X-ray spectral study of 24 type 1
active galactic nuclei, MNRAS, 286, 513

\noindent Risaliti, G. et al. 1999, The Distribution of Absorbing
Column Densities among Seyfert 2 Galaxies, ApJ, 522, 157

\noindent Rutledge, R.E. \& Sako, M. 2003, Statistical re-examination
of reported emission lines in the X-ray afterglow of GRB 011211,
MNRAS, 339, 600

\noindent Sakelliou, I. et al. 2002, High Resolution Soft X-ray
Spectroscopy of M87 with the Reflection Grating Spectrometers on
{\it XMM-Newton}, A\&A, 391, 903

\noindent Sako, M. et al. 2001, Complex resonance absorption
structure in the X-ray spectrum of IRAS 13349+2438, A\&A, 365, L168

\noindent Sambruna, R., Eracleous, M., Mushotzky, R.F., 1999, An X-Ray
Spectral Survey of Radio-loud Active Galactic Nuclei with ASCA, ApJ,
526, 60

\noindent Sarazin, C.L., 1986, X-ray emission from clusters of
galaxies, Rev Mod Phys, 58, 1

\noindent Sasaki, M., et al.\ 2000, {\it ROSAT} HRI
catalogue of X-ray sources in the LMC region, A\&AS 143, 391

\noindent Sasaki, M., et al.\ 2003, {\it
XMM-Newton} observations of High Mass X-ray Binaries in the SMC, A\&A,
403, 901 

\noindent Schenker, K. et al.\ 2002, AE Aquarii: how cataclysmic
variables descend from supersoft binaries, MNRAS 337, 1105

\noindent Severgnini, P. et al. 2003, {\it XMM-Newton} observations expose
AGN in apparently normal galaxies, A\&A, in the press
(astro-ph/0304308)

\noindent Shu, F.H. et al.\ 1994, Magnetocentrifugally driven flows
from young stars and disks. I. A generalized model, ApJ 429, 781

\noindent Sidoli, L. et al.\ 2001, An {\it XMM-Newton} study of the
X-ray binary MXB 1659$-$298 and the discovery of narrow X-ray
absorption lines, A\&A 379, 540

\noindent Sion, E.M. 1999, White Dwarfs in Cataclysmic Variables, PASP
111, 532

\noindent Skinner, S.L., et al.\ 2002, {\it XMM-Newton} detection of
hard X-ray emission in the Nitrogen-type Wolf-Rayet star WR110, ApJ
579, 76

\noindent Smith, D.A. \& Done, C. 1996,Unified theories of active
galactic nuclei: a hard X-ray sample of Seyfert 2 galaxies, MNRAS,
280, 355

\noindent Stelzer, B., \& Burwitz, V. 2003, Castor A and Castor B
resolved in a simultaneous {\it Chandra} and {\it XMM-Newton}
observation, A\&A 402, 719

\noindent Stelzer, B. et al.\ 2002, Simultaneous X-ray spectroscopy of
YY Gem with {\it Chandra} and {\it XMM-Newton}, A\&A 392, 585

\noindent Strohmayer, T.E., \& Mushotzky, R.F. 2003, Discovery of
X-Ray quasi-periodic oscillations from an ultraluminous X-Ray source
in M82: Evidence against beaming, ApJ 552, L109

\noindent Str\"uder, L. et al., 2001, The European Photon Imaging
Camera on {\it XMM-Newton}: The pn-CCD camera, A\&A, 365, L5

\noindent Supper, R. et al.\ 2001, The second {\it ROSAT} PSPC survey
of M 31 and the complete {\it ROSAT} PSPC source list, A\&A 373, 63

\noindent Sutaria, F.K. et al.\ 2002, {\it XMM-Newton} detection of
Nova Muscae 1991 in quiescence, A\&A 391, 993

\noindent Tamura, T. et al. 2001a, {\it XMM-Newton} observations of the
cluster of galaxies Abell 496. Measurements of the elemental
abundances in the intracluster medium, A\&A, 379, 107

\noindent Tamura, T. et al. 2001b, X-ray spectroscopy of the cluster
of galaxies Abell 1795 with {\it XMM-Newton}, A\&A, 365, L87

\noindent Tanaka, Y. et al. 1995, Gravitationally Redshifted Emission
Implying an Accretion Disk and Massive Black-Hole in the Active Galaxy
MCG-6-30-15, Nat, 375, 659

\noindent Tanaka, Y., \& Shibazaki, N. 1996, X-ray novae, ARA\&A 34,
607

\noindent Tiengo, A. et al.\ 2002, The anomalous X-ray pulsar 1E
1048.1$-$5937: Phase resolved spectroscopy with the {\it XMM-Newton}
satellite, A\&A 383, 182

\noindent Topka, K. et al.\ 1979, Detection of soft X-rays from Alpha
Lyrae and Eta Bootis with an imaging X-ray telescope, ApJ, 229, 661

\noindent Trudolyubov, S.P., et al.\ 2001, Bright X-Ray transients in
the Andromeda Galaxy observed with {\it Chandra} and {\it XMM-Newton},
ApJ 563, L119

\noindent Trudolyubov, S.P. et al.\ 2002a, On the X-Ray source
luminosity distributions in the bulge and disk of M31: First results
from the {\it XMM-Newton} survey, ApJ 571, L17

\noindent Trudolyubov, S.P. et al.\ 2002b, The discovery of a 2.78
hour periodic modulation of the X-Ray flux from Globular Cluster
Source Bo 158 in M31, ApJ 581, L27

\noindent Turner, M.J.L. et al. 2001, The European Photon Imaging
Camera on {\it XMM-Newton}: The MOS cameras : The MOS cameras, A\&A, 365,
L27

\noindent Vaiana G.S., et al.\ 1981, Results from an extensive {\it
Einstein} stellar survey, ApJ 244, 163

\noindent Vilhu, O. 1984, The nature of magnetic activity in lower
main sequence stars, A\&A 133, 117

\noindent Warwick, R.S. et al.\ 2001, The extended X-ray halo of the
Crab-like SNR G21.5$-$0.9, A\&A 365, L248

\noindent Watson, M.G. et al. 2001,The {\it XMM-Newton} Serendipitous
Survey. I. The role of {\it XMM-Newton} Survey Science Centre. A\&A, 365,
L51
  
\noindent White, S.D.M., et al., 1993, X-ray archaeology in the Coma
cluster, MNRAS, 261, L8

\noindent Wijnands, R., \& Van der Klis, M. 1998, A millisecond pulsar
in an X-ray binary system, Nat. 394, 344

\noindent Wijnands, R. 2003, An {\it XMM-Newton} observation during
the 2000 outburst of SAX J1808.4$-$3658, ApJ 588, 425

\noindent Wilkes, B.J. \& Elvis, M.S. 1987, Quasar energy
distributions. I - Soft X-ray spectra of quasars, ApJ, 323, 243

\noindent Willingale, R. et al.\ 2002, X-ray spectral imaging and
Doppler mapping of Cassiopeia A, A\&A 381, 1039

\noindent Willingale, R. et al.\ 2003, The mass and energy budget of
Cassiopeia A, A\&A 398, 1021

\noindent Wilms, J. et al.  2001, XMM-EPIC observation of
MCG-6-30-15: Direct evidence for the extraction of energy from a
spinning black hole?, MNRAS, 328, L27

\noindent Worrall, D.M. \& Birkinshaw, M., 2003, The temperature and
distribution of gas in CL 0016+16 measured with {\it XMM-Newton}, MNRAS, in
the press (astro-ph/0301123)

\end{document}